\documentclass[11pt]{article}
\usepackage{amsmath, amssymb, amsfonts, amsthm}

\newtheorem{theorem}{Theorem}[section]
\newtheorem{proposition}[theorem]{Proposition}

\newtheorem{lemma}[theorem]{Lemma}

\newcommand{\rd}{{\rm d}}
\newcommand{\be}{\begin{equation}}
\newcommand{\ee}{\end{equation}}
\newcommand{\bey}{\begin{eqnarray}}
\newcommand{\eey}{\end{eqnarray}}

\newcommand{\eps}{\varepsilon}

\newcommand{\bx}{{\bf x}}

\newcommand{\com}[2]{ \left [ #1 \, , #2 \right ] }

\newcommand{\ph}{\varphi}

\newcommand{\balpha}{{\boldsymbol \alpha}}

\newcommand{\cU}{{\cal U}}

\newcommand{\bR}{{\mathbb R}}

\newcommand{\bN}{{\mathbb N}}

\newcommand{\tr}{\mbox{Tr}}

\newcommand{\wt}{\widetilde}
\newcommand{\wh}{\widehat}

\newcommand{\const}{\mathrm{const}}

\newcommand{\cS}{{\cal S}}

\newcommand{\cA}{{\cal A}}

\newcommand{\cE}{{\cal E}}

\newcommand{\cK}{{\cal K}}
\newcommand{\cH}{{\cal H}}
\newcommand{\cL}{{\cal L}}

\input epsf

\newcommand{\donothing}[1]{}

\begin{document}
\title{Mean Field Dynamics of Boson Stars}
\author{Alexander Elgart and Benjamin
Schlein\thanks{Supported by NSF Postdoctoral
Fellowship.} \\
\\
Department of Mathematics, Stanford University\\ Stanford, CA
94305, USA}

\maketitle

\begin{abstract}
We consider a quantum mechanical system of $N$ bosons with
relativistic dispersion interacting through a mean field Coulomb
potential (attractive or repulsive). We choose the initial wave
function to describe a condensate, where the $N$ bosons are all in
the same one-particle state. Starting from the $N$-body
Schr\"odinger equation, we prove that, in the limit $N \to
\infty$, the time evolution of the one-particle density is
governed by the relativistic nonlinear Hartree equation. This
equation is used to describe the dynamics of boson stars
(Chandrasekhar theory). The corresponding static problem was
rigorously solved in \cite{LY}.
\end{abstract}

\section{Introduction}
\label{sec:intro} \setcounter{equation}{0}

We consider a system of $N$ identical bosons with relativistic
dispersion relation and with a mean field Coulomb interaction. The
system is described on the Hilbert space $L_s^2 (\bR^{3N}, \rd
\bx)$, the subspace of $L^2 (\bR^{3N}, \rd \bx)$ containing all
functions symmetric with respect to permutations. The Hamiltonian
of the system is given by
\begin{equation}\label{eq:ham}
H_N = \sum_{j=1}^N (1-\Delta_j)^{1/2} + \frac{\lambda}{N}
\sum_{i<j}^N \frac{1}{|x_i - x_j|} \, .
\end{equation}
Here we use units with $\hbar = m = 1$ (where $m$ denotes the mass
of the bosons). The Hamiltonian $H_N$ defines a mean field
interaction among the bosons because the coupling constant is
proportional to $1/N$: with this scaling the kinetic and potential
part of the energy are typically of the same order. This condition
is necessary in order to have a mean field description of the
system in the limit of large $N$.

The constant $\lambda$ can be positive or negative, corresponding
to repulsive and attractive interaction. In the case of repulsive
interaction we will have no restriction on the value of $\lambda$.
The choice of a negative $\lambda$, which corresponds to an
attractive Coulomb potential, leads to a Chandrasekhar theory of
boson stars, where general relativity effects are neglected. In
this case we need to impose the condition $\lambda > - 4/\pi$. To
understand why this assumption is necessary, note that, if
$\lambda >-4/\pi$, the Hamiltonian $H_N$, with domain $D(H_N) =
H^1(\bR^{3N})$, is self adjoint and stable, in the sense that the
ground state energy of $H_N$ divided by the number of particle
$N$, is bounded below, uniformly in $N$. On the other hand, for
$\lambda < -4/\pi$, the Hamiltonian $H_N$ is unstable: the ground
state energy per particle diverges to $-\infty$ as $N \to \infty$.
This phenomenon is usually referred to as the collapse of the
system: the energy is minimized by letting the particles closer
and closer, because the increase of the kinetic energy (due to the
localization of the particles) is not enough, if $\lambda <
-4/\pi$, to compensate for the decrease of the potential energy.
These results were derived in \cite{LY}, where the authors prove
that the energy per particle of the ground state of the system is
given, in the limit $N \to \infty$, by the minimization of a
one-particle energy functional:
\begin{equation}\label{eq:GS}
\begin{split}
\lim_{N \to \infty} \frac{E_{\text{GS}} (\lambda)}{N} &=
\inf_{\ph: \| \ph \|=1} \, \cE (\ph, \overline{\ph}) \\ &=
\inf_{\ph: \| \ph \|=1} \left( \int \rd x \, |(1-\Delta)^{1/4} \ph
(x)|^2 + \frac{\lambda}{2} \int \rd x \, (V
* |\ph|^2) (x) \, |\ph (x)|^2 \right) \, .
\end{split}
\end{equation}
Here we defined $V(x) = |x|^{-1}$. The r.h.s. is negative infinity
if $\lambda < -4/\pi$. The existence of a critical coupling
$\lambda_{\text{crit}} = -4/\pi$ is due to the fact that the
kinetic energy, which behaves, for large momenta, as $|\nabla|$,
and the potential $|x|^{-1}$ scale in the same way.

In this paper we are interested in the dynamics generated by $H_N$
in the limit $N \to \infty$. {F}rom (\ref{eq:GS}) one can expect
that the macroscopic dynamics of the system are described, in the
limit $N \to \infty$, by the one particle nonlinear relativistic
Hartree equation
\begin{equation}\label{eq:hartree}
i\partial_t \ph_t (x) = \frac{\delta \cE}{\delta \overline{\ph}}
(\ph_t (x) ,\overline{\ph_t} (x)) = (1-\Delta)^{1/2} \ph_t (x) +
\lambda (V * |\ph_t|^2) (x) \ph_t (x) \, .
\end{equation}
Some important properties of this equation, such as the global
well-posedness for $\lambda > \lambda_{\text{crit}}$, are proven
in \cite{L}.

To formulate the convergence towards the nonlinear Hartree
equation (\ref{eq:hartree}) more precisely, we define next the
marginal distributions of an $N$-particle wave function $\psi_N$,
and we investigate their time evolution. Given an $N$-particle
wave function $\psi (\bx) \in L^2_s (\bR^{3N}, \rd\bx)$ with $\|
\psi \| =1$, we define the corresponding density matrix
$\gamma_{N} = |\psi_N \rangle \langle \psi_N|$ as the orthogonal
projection onto $\psi$. The kernel of $\gamma_N$ is
\begin{equation*}
\gamma_N (\bx; \bx') = \psi_N (\bx) \overline{\psi_N} (\bx') \, .
\end{equation*}
More generally, a density matrix $\gamma_N$ is a non-negative
trace class operator on $L_s^2 (\bR^{3N})$ with $\tr \, \gamma_N
=1$. For $k=1, \dots N$, the $k$-particle marginal distribution of
$\gamma_N$, denoted by $\gamma^{(k)}_N$, is defined by taking the
partial trace over the last $N-k$ variables of $\gamma_N$, that
is, the kernel of $\gamma_N^{(k)}$ is given by
\begin{equation*}
\gamma^{(k)}_N (\bx_k ; \bx'_k) = \int \rd \bx_{N-k} \, \gamma_N
(\bx_k , \bx_{N-k} ; \bx'_k , \bx_{N-k}) \, .
\end{equation*}
Here and henceforth we use the notation \[ \bx = (x_1, \dots ,x_N)
, \quad \bx_k = (x_1, \dots , x_k), \quad \bx_{N-k} = (x_{k+1} ,
x_{k+2}, \dots , x_N) \] with $x_j \in \bR^3$, for all $j=1, \dots
N$, and analogously for the primed variables. Since $\tr \,
\gamma_N =1$, it follows immediately that $\tr \, \gamma_N^{(k)}
=1$ for all $k =1, \dots ,N$.

The time evolution of the system is governed by the Schr\"odinger
equation
\begin{equation}\label{eq:schr}
i \partial_t \psi_{N,t} = H_N \, \psi_{N,t} \,
\end{equation}
or, equivalently, by the Heisenberg equation
\begin{equation*}
i\partial_t \gamma_{N,t} = [ H_N, \gamma_{N,t}]
\end{equation*}
for the dynamics of $\gamma_N$. {F}rom the Schr\"odinger equation
we can also derive a hierarchy of $N$ equations, usually called
the BBGKY hierarchy, describing the evolution of the marginal
distributions $\gamma_{N,t}^{(k)}$ for $k=1,\dots ,N$. Using the
permutation symmetry we find, for $k=1,\dots,N$,
\begin{equation}\label{eq:BBGKY}
\begin{split}
i\partial_t \gamma^{(k)}_{N,t}  = \; &\sum_{j=1}^k \,
\left[(1-\Delta_{x_j})^{1/2}, \gamma_{N,t}^{(k)} \right] +
\frac{\lambda}{N} \sum_{1 \leq i<j \leq k} \left[ V(x_i - x_j),
\gamma^{(k)}_{N,t}\right] \\ &+\lambda (1 -\frac{k}{N})
\sum_{j=1}^k \tr_{k+1} \left[ V(x_j -x_{k+1}),
\gamma_{N,t}^{(k+1)}\right] \, .
\end{split}
\end{equation}
(Recall that $V(x) = |x|^{-1}$). Here we use the convention
$\gamma^{(k)}_N =0$, if $k = N+1$. Moreover we denote by
$\tr_{k+1}$ the partial trace over the variable $x_{k+1}$. In
terms of the kernels $\gamma^{(k)}_{N,t} (\bx_k,\bx'_k)$ the last
equation can be written as
\begin{equation*}
\begin{split}
i\partial_t \gamma^{(k)}_{N,t} &(\bx_k;\bx'_k) = \sum_{j=1}^k \,
((1-\Delta_{x_j})^{1/2} - (1- \Delta_{x'_j})^{1/2})
\gamma_{N,t}^{(k)} (\bx_k;\bx'_k)
\\ &+ \frac{\lambda}{N} \sum_{1 \leq i<j \leq k} \left( V(x_i - x_j) -
V(x'_i -x'_j) \right) \gamma^{(k)}_{N,t} (\bx_k;\bx'_k) \\
&+\lambda (1 -\frac{k}{N}) \sum_{j=1}^k \int \rd x_{k+1} \left(
V(x_j -x_{k+1}) - V(x'_j -x_{k+1}) \right) \gamma_{N,t}^{(k+1)}
(\bx_k, x_{k+1} ; \bx'_k , x_{k+1}) \, .
\end{split}
\end{equation*}
In the limit $N \to \infty$ the BBGKY hierarchy (\ref{eq:BBGKY})
formally converges to the infinite hierarchy of equations
\begin{equation}\label{eq:BBGKYinf}
\begin{split}
i\partial_t \gamma^{(k)}_{t} = \, &\sum_{j=1}^k \, \left[
(1-\Delta_{x_j})^{1/2}, \gamma_{t}^{(k)}\right] + \lambda
\sum_{j=1}^k  \tr_{k+1} \left[ V(x_j -x_{k+1}),
\gamma_{t}^{(k+1)}\right]
\end{split}
\end{equation}
for all $k \geq 1$. It is easy to check that, if the initial data
is factorized, that is if
\begin{equation*}
\gamma^{(k)}_{t=0} (\bx_k ;\bx'_k) = \prod_{j=1}^k \ph (x_j)
\overline{\ph} (x'_j)
\end{equation*}
for all $k \geq 1$, then the infinite hierarchy
(\ref{eq:BBGKYinf}) has the solution
\[ \gamma^{(k)}_t (\bx_k; \bx'_k) = \prod_{j=1}^k \ph_t (x_j)
\overline{\ph_t} (x'_j) \] where $\ph_t (x)$ is the solution of
the non-linear one-particle Hartree equation (\ref{eq:hartree})
with initial data $\ph_{t=0} = \ph$ (for $\lambda >
\lambda_{\text{crit}} = -4/\pi$, (\ref{eq:hartree}) is known to
have a unique global solution in the space $H^{m/2} (\bR^3)$, for
every $m\geq 1$, see \cite{L}). Hence, if we consider a factorized
initial wave function $\psi_N (\bx) = \prod_{j=1}^N \ph (x_j)$, if
we fix $k \geq 1$ and $t \in \bR$, and if we denote by
$\gamma^{(k)}_{N,t}$ the $k$-particle marginal distribution
associated with the time evolution $\psi_{N,t}$ of $\psi_N$, then
we can expect that, in a suitable weak topology,
\begin{equation}\label{eq:limitgamma}
\gamma^{(k)}_{N,t} (\bx_k;\bx'_k) \to \gamma^{(k)}_{t}
(\bx_k;\bx'_k) = \prod_{j=1}^k \ph_t (x_j) \overline{\ph_t} (x'_j)
\quad \text{for } N \to \infty\,
\end{equation}
where $\ph_t$ is the solution of the Hartree equation
(\ref{eq:hartree}). The aim of this paper is to give a rigorous
proof of this statement.

\bigskip

The first result of this type was proven in \cite{Hepp} for
non-relativistic dispersions and for smooth potentials. This work
was generalized to bounded potentials in \cite{Sp}. In \cite{GV}
the authors show the convergence (\ref{eq:limitgamma}), for a
non-relativistic dispersion, and for any integrable potential:
they use the formalism of second quantization and they need the
initial state to be coherent (and thus the number of particles
cannot be fixed). In \cite{EY} the convergence
(\ref{eq:limitgamma}) was proven for bosons with non-relativistic
dispersion interacting through a Coulomb potential. Partial
results for the non-relativistic Coulomb case were also
established in \cite{BGM}. In \cite{EESY}, a joint work with L.
Erd\H os and H.-T. Yau, we consider a system of $N$
non-relativistic bosons, interacting, in the mean-field scaling,
through an $N$-dependent potential $V_N (x) = N^{3\beta}
V(N^{\beta} x)$, with $0 < \beta < 3/5$, which converges to a
delta-function in the limit $N \to \infty$: for this potential we
prove the convergence of solutions of the finite hierarchy
(\ref{eq:BBGKY}) (with $V(x)$ replaced by $V_N(x)$) to solutions
of the infinite hierarchy (\ref{eq:BBGKYinf}) (with $V$ replaced
by $\delta (x)$). A recent overview of rigorous results and open
problems concerning the nonlinear Hartree equation and its
derivation as the mean field limit of large bosonic systems can be
found in \cite{FL}.

Note that the non-relativistic case considered in all these works
is technically easier than the relativistic case we are
considering in the present paper: the presence of a quadratic
dispersion (the non-relativistic kinetic energy of a boson is
given by the negative Laplacian, quadratic in the momentum) makes
the control of the Coulomb singularity much simpler than in the
relativistic case, where the kinetic energy is only linear in the
momentum (for large momenta, $(1+p^2)^{1/2} \simeq |p|$). Hence,
although our general strategy is the same as in \cite{EY}, we need
here more refined estimates.

\bigskip

Next we explain the methods we use to prove (\ref{eq:limitgamma})
in some details. The main technical tool in our analysis is an
a-priori estimate (Theorem \ref{thm:apriori}), which guarantees a
certain smoothness of the $N$-body wave function (and thus of the
corresponding marginal densities), uniformly in $N$ and $t$. As in
\cite{EY}, we derive the a-priori estimate from energy estimates,
which control powers of the energy in terms of the corresponding
powers of the kinetic energy. In order to prove these energy
estimates we need to introduce an $N$-dependent cutoff in the
Hamiltonian. For $\eps >0$ we define the cutoff Hamiltonian
\begin{equation}\label{eq:cutoffham}
\wt H_N = \sum_{j=1}^N (1 - \Delta_j)^{1/2} + \lambda \sum_{i<j}^N
\frac{1}{|x_i -x_j| + \eps N^{-1}}\,.
\end{equation}
In $\wt H_N$ the Coulomb singularity has been regularized on the
length scale $|x| \simeq \eps N^{-1}$. In
Proposition~\ref{prop:lower} we then prove the following operator
bound (valid on the space $L_s^2 (\bR^{3N})$ of functions
invariant w.r.t. permutations) for powers of the Hamiltonian $\wt
H_N$:
\begin{equation}\label{eq:low}
\wt H_N^k \geq \; C^k N^k \; (1-\Delta_1)^{1/2} (1-\Delta_2)^{1/2}
\dots (1-\Delta_k)^{1/2}
\end{equation}
for all $k \geq 1$, and for all $N$ large enough (depending on $k$
and on the cutoff $\eps$). Hence, if the initial state $\psi_N \in
L^2_s (\bR^{3N})$ is such that
\begin{equation}\label{eq:expect}
(\psi_N, \wt H^k_N \psi_N) \leq C^k N^k
\end{equation}
for all $k \geq 0$, then, using the conservation of the energy,
from (\ref{eq:low}) we can derive bounds on higher derivatives of
the solution $\psi^{\eps}_{N,t}$ of the Schr\"odinger equation
\begin{equation}\label{eq:schrcut}
i\partial_t \psi_{N,t}^{\eps} = \wt H_N \psi_{N,t}^{\eps}
\end{equation}
with regularized interaction, and with initial data
$\psi_{N,t=0}^{\eps} = \psi_N$. Unfortunately, a typical
factorized wave function $\psi_N (\bx) = \prod_{j=1}^N \ph (x_j)$
does not satisfies (\ref{eq:expect}) (unless $\ph \in C_0^{\infty}
(\bR^3)$, an assumption we want to avoid). Therefore, the
introduction of an additional cutoff in the initial one-particle
wave function $\ph$ is required. Given $\kappa>0$ and a one
particle wave function $\ph \in H^1 (\bR^3)$, we define its
regularized version by \[ \ph^{\kappa} = \exp (-\kappa |p|/N) \ph
\quad \quad \text{with } p = -i\nabla \, . \] Then the regularized
$N$-particles wave function $\psi_{N}^{\kappa} (\bx) =
\prod_{j=1}^N \ph^{\kappa} (x_j)$ can be proven to satisfy
(\ref{eq:expect}) for every $k \geq 1$ and for all $N$ large
enough (depending on $k$ and $\kappa$). This follows from
Proposition \ref{prop:upper}, where we derive an upper bound for
the expectation of powers of the Hamiltonian. Hence if we denote
by $\psi_{N,t}^{\eps,\kappa}$ the evolution of the regularized
initial data $\psi_N^{\kappa}$ generated by $\wt H_N$, then
(\ref{eq:low}) implies that
\begin{equation}\label{eq:apr}
(\psi_{N,t}^{\eps, \kappa}, (1 -\Delta_1)^{1/2} \dots
(1-\Delta_k)^{1/2} \psi_{N,t}^{\eps, \kappa}) \leq C^k
\end{equation}
for all $k \geq 1$ and for all $N$ large enough (depending on $k$,
$\eps$ and $\kappa$). For fixed $\eps, \kappa >0$ we denote by
$\wt \Gamma_{N,t} = \{ \wt \gamma^{(k)}_{N,t} \}_{k=1}^N$ the
family of marginal distributions associated with the wave function
$\psi_{N,t}^{\eps,\kappa}$.

Using the bound (\ref{eq:apr}) we can prove the compactness of the
sequence $\wt \Gamma_{N,t}$ with respect to an appropriate weak
topology. Moreover we can show that any limit point
$\Gamma_{\infty,t} = \{ \gamma^{(k)}_{\infty,t} \}_{k \geq 1}$ of
$\wt \Gamma_{N,t}$ satisfies the infinite hierarchy
(\ref{eq:BBGKYinf}) with initial value $\gamma^{(k)}_{\infty,t=0}
(\bx_k;\bx'_k) = \prod_{j=1}^k \ph (x_j) \overline{\ph} (x'_j)$
(the cutoffs $\eps$ and $\kappa$, which regularize the interaction
and the initial data, disappear in the limit $N \to \infty$) and
that the solution of (\ref{eq:BBGKYinf}) is unique (in an
appropriate space). This implies that
\[ \gamma^{(k)}_{\infty,t} (\bx_k ; \bx'_k) = \prod_{j=1}^k \ph_t
(x_j) \overline{\ph_t} (x_j),\] for every $k \geq 1$, where
$\ph_t$ is the solution of the non-linear Hartree equation
(\ref{eq:hartree}) with initial data $\ph_{t=0} =\ph$. Finally, we
prove that the difference between the physical evolution
(generated by the Hamiltonian $H_N$) of the initial wave function
$\psi_N (\bx) = \prod_{j=1}^k \ph (x_j)$ and the modified
evolution (generated by $\wt H_N$) of the approximate initial wave
function $\psi_{N}^{\kappa}$ converges to zero, for $\eps$ and
$\kappa$ converging to zero, uniformly in $N$ (for $N$ large
enough). This concludes the proof of (\ref{eq:limitgamma}).

\bigskip

The paper is organized as follows. In Section \ref{sec:topo} we
introduce some Banach spaces of density matrices, and we equip
them with weak topologies, useful to take the limit $N \to
\infty$. In Section \ref{sec:main} we state our main result and
give a sketch of its proof. In Section \ref{sec:apriori} we prove
the energy estimates for the regularized Hamiltonian $\wt H_N$,
and we use them to derive the a-priori bound (\ref{eq:apr}). In
Section~\ref{sec:compact} we show the compactness of the sequence
of marginals $\wt \gamma^{(k)}_{N,t}$ associated with the wave
function $\psi_{N,t}^{\eps,\kappa}$ with respect to the weak
topology. In Section \ref{sec:conv} we demonstrate that any limit
point of the marginal densities $\wt \Gamma_{N,t} = \{\wt
\gamma^{(k)}_{N,t}\}_{k\geq 1}$ satisfies the infinite hierarchy
(\ref{eq:BBGKYinf}). In Section \ref{sec:unique} we show the
uniqueness of the solution of (\ref{eq:BBGKYinf}). In Section
\ref{sec:comp} we remove the cutoffs, letting $\eps, \kappa \to
0$. Finally, in Section \ref{sec:techn}, we collect some important
technical lemmas, used throughout the paper.

\bigskip

\emph{Notations.} Throughout the paper, we will use the notation
$S_j = (1 + p_j^2)^{1/4} = (1- \Delta_j)^{1/4}$ for integer $j
\geq 1$. Moreover, with an abuse of notation, we will denote by
$\| \ph \|$ the $L^2$-norm of the function $\ph$ and by $\| A \|$
the operator norm of the the operator $A$. We will use the symbol
$\tr_k$ to denote the partial trace over the $k$'th variable. In
general we will use the symbol $\tr$ to denote the trace over all
variables involved: sometimes, instead of $\tr$, we will use the
symbol $\tr^{(k)}$ to stress the total number of variables over
which the trace is taken. In general, we will denote by $C$ a
universal constant, depending, possibly, only on the coupling
constant $\lambda$ and on the $H^1$-norm of the initial
one-particle wave function $\ph$. If $C$ also depends on other
quantities, we will usually stress them explicitly.

\bigskip

\emph{Acknowledgements.} We are grateful to H.-T. Yau and L. Erd\H
os for suggesting this problem to us, and for very useful
discussions. We thank G. M. Graf for helpful suggestions. B. S.
would also like to thank J. Fr\"ohlich, S. Schwarz and E. Lenzmann
for stimulating discussions related to the relativistic Hartree
equation.

\section{Banach Spaces of Density Matrices}
\label{sec:topo} \setcounter{equation}{0}

For $k \geq 1$, we denote by $\cL_k^1$ and by $\cK_k$ the space of
trace class and, respectively, of compact operators on the
$k$-particle Hilbert space $L^2 (\bR^{3k}, \rd \bx_k)$. It is a
well known fact (see, for example, Theorem VI.26 in \cite{RS})
that $(\cL^1_k, \| \, . \, \|_1) = (\cK_k , \| \, . \, \|)^*$,
where $\| \, . \, \|_1$ denotes the trace norm, and $\|\, .\, \|$
the operator norm. This induces a weak* topology on $\cL^1_k$. We
define $\cL^1$ as the direct sum over $k \geq 1$ of the spaces
$\cL_k^1$, that is
\begin{equation*}
\cL^1 = \bigoplus_{k \geq 1} \cL^1_k = \{ \Gamma = \{ \gamma^{(k)}
\}_{k \geq 1} : \gamma^{(k)} \in \cL_k^1, \quad \forall \; k \geq
1 \}\,.
\end{equation*}
We equip the space $\cL^1$ with the product of the weak*
topologies on $\cL^1_k$ for all $k \geq 1$. A sequence $\Gamma_N =
\{ \gamma^{(k)}_N \}_{k \geq 1}$ converges to $\Gamma = \{
\gamma^{(k)}\}_{k \geq 1}$ in $\cL^1$ for $N \to \infty$ if and
only if $\gamma_N^{(k)}$ converges to $\gamma^{(k)}$ with respect
to the weak* topology of $\cL^1_k$ for all $k \geq 1$.

\bigskip

To prove our main theorem (Theorem \ref{thm:main}), we also need
to define a different topology, which makes use of the smoothness
of the marginal densities $\gamma_{N,t}^{(k)}$ and allows us to
pass the smoothness to the limit $N \to \infty$. For $\gamma^{(k)}
\in \cL^1_k$, we define the norm
\begin{equation*}
\| \gamma^{(k)} \|_{\cH_k} = \tr \,| S_1 \dots S_k \,\gamma^{(k)}
S_k \dots S_1|
\end{equation*}
and the corresponding Banach space \[ \cH_k = \{ \gamma^{(k)} \in
\cL^1_k : \| \gamma^{(k)} \|_{\cH_k} < \infty \}. \] Recall that
we use the notation $S_j = (1 + p_j^2)^{1/4} = (1 -
\Delta_j)^{1/4}$. Moreover we define the space of operators
\[ \cA_k = \{ T^{(k)} = S_1 \dots S_k K^{(k)} S_k \dots S_1 : K^{(k)} \in
\cK_k \} \] with the norm
\[ \| T^{(k)} \|_{\cA_k} = \| S_1^{-1} \dots S_k^{-1} T^{(k)}
S_k^{-1} \dots S_1^{-1} \| \] where $\| \, . \, \|$ denotes the
operator norm. Then we have, for every $k \geq 1$,
\begin{equation}\label{eq:dual}
(\cH_k , \| \, . \, \|_{\cH_k}) = ( \cA_k, \| \, . \,
\|_{\cA_k})^* \end{equation} and thus we have a weak* topology on
the space $\cH_{k}$. The proof of (\ref{eq:dual}) is identical to
the proof of Lemma 3.1 in \cite{EY}. Note that the weak* topology
on $\cH_{k}$ is stronger than the weak* topology on the larger
space $\cL^1_k$: if a sequence $\gamma_N^{(k)}$ converges to
$\gamma^{(k)}$ with respect to the weak* topology of $\cH_k$, then
it also converges with respect to the weak* topology of $\cL^1_k$.

\bigskip

We will prove that the densities $\wt \gamma_{N,t}^{(k)}$
associated with the wave function $\psi_{N,t}^{\eps,\kappa}$
(defined by the regularized Schr\"odinger equation
(\ref{eq:schrcut})), are continuous functions of $t$ with values
in $\cH_k$ (continuous w.r.t. the weak* topology of $\cH_k$). In
order to prove the continuity of limits of $\gamma^{(k)}_{N,t}$,
we want to invoke the Arzela-Ascoli Theorem, which is only
applicable to metric spaces. Fortunately, since the space $\cA_k$
is separable (which follows from the fact that the space $\cK_k$
of compact operators is separable), it turns out that the weak*
topology of $\cH_k$, when restricted to the unit ball of $\cH_k$,
is metrizable. In fact, because of the separability of $\cA_k$, we
can fix a dense countable subset of the unit ball of $\cA_k$: we
denote it by $\{J^{(k)}_i\}_{i \ge 1} \in \cA_k$, with $\|
J^{(k)}_i \|_{\cA_k} =1$ for all $i \ge 1$. Using the operators
$J^{(k)}_i$ we define the following metric on $\cH_k$: for
$\gamma^{(k)}, \wt \gamma^{(k)} \in \cH_k$ we set
\begin{equation}\label{eq:rho}
\rho_k (\gamma^{(k)}, \wt \gamma^{(k)}) : = \sum_{i=1}^\infty
2^{-i} \left| \tr \; J^{(k)}_i \left( \gamma^{(k)} - \wt
\gamma^{(k)} \right) \right| \, .
\end{equation}
Then the topology induced by the metric $\rho_k (.,.)$ and the
weak* topology are equivalent on the unit ball of $\cH_k$ (see
\cite{Ru}, Theorem 3.16). In other words, a uniformly bounded
sequence $\gamma_N^{(k)} \in \cH_k$ converges to $\gamma^{(k)} \in
\cH_k$ with respect to the weak* topology, if and only if $\rho_k
(\gamma^{(k)}_N , \gamma^{(k)}) \to 0$ for $N \to \infty$.

\bigskip

For a fixed $T \geq 0$ we consider the space $C ([0,T], \cH_k)$ of
functions of $t \in [0,T]$ with values in $\cH_k$, which are
continuous with respect to the metric $\rho_k$ (for uniformly
bounded functions, the continuity with respect to $\rho_k$ is
equivalent to continuity with respect to the weak* topology of
$\cH_k$). On the space $C ([0,T], \cH_k)$ we define the metric
$\widehat \rho_k$ by
\begin{equation}\label{eq:hatrho}
\widehat \rho_k (\gamma^{(k)} (.) , \wt \gamma^{(k)} (.)) =
\sup_{t \in [0,T]} \rho_k (\gamma^{(k)} (t) , \wt \gamma^{(k)}
(t))\,.
\end{equation}

Finally, we define the space $\cH$ as the direct sum over $k \geq
1$ of the spaces $\cH_k$, that is
\begin{equation*}
\cH = \bigoplus_{k \geq 1} \cH_k = \{ \Gamma = \{ \gamma^{(k)}
\}_{k \geq 1} : \gamma^{(k)} \in \cH_k, \quad \forall \, k \geq 1
\} \, ,
\end{equation*}
and, for a fixed $T \geq 0$, we consider the space \[ C([0,T] ,
\cH) = \bigoplus_{k \geq 1} C([0,T], \cH_k),\] equipped with the
product of the topologies induced by the metric $\widehat
\rho^{(k)}$ on $C([0,T],\cH_k)$. That is, for $\Gamma_{N,t}= \{
\gamma_{N,t}^{(k)} \}_{k \geq 1}$ and $ \Gamma_t = \{
\gamma^{(k)}_t \}_{k \geq 1}$ in $C ([0,T], \cH)$, we have
$\Gamma_{N,t} \to \Gamma_t$ for $N \to \infty$ if and only if, for
every fixed $k \geq 1$,
\begin{equation*}
\widehat \rho_k (\gamma_{N,t}^{(k)},\gamma_t^{(k)}) = \sup_{t \in
[0,T]} \rho_k (\gamma_{N,t}^{(k)}, \gamma_t^{(k)}) \to 0
\end{equation*}
for $N \to \infty$.

\section{Main Result}
\label{sec:main}\setcounter{equation}{0}

We are now ready to formulate our main theorem. We also give its
proof, using results which will be established later on.

\begin{theorem}\label{thm:main}
Let $\ph \in H^1 (\bR^3)$, with $\| \ph \| =1$, and $\psi_{N}
(\bx) = \prod_{j=1}^N \ph (x_j)$. Denote by $\,\psi_{N,t}$ the
time evolution of $\psi_{N}$ determined by the Schr\"odinger
equation (\ref{eq:schr}), and let $\gamma^{(k)}_{N,t}$, for
$k=1,\dots,N$, be the marginal distributions corresponding to
$\psi_{N,t}$. Assume that the coupling constant $\lambda >
-4/\pi$. Then for all fixed $t \in \bR$, we have, for $N \to
\infty$,
\begin{equation}\label{eq:conve}
\Gamma_{N,t} = \{ \gamma_{N,t}^{(k)} \}_{k \geq 1} \to \Gamma_{t}
= \{ \gamma_{t}^{(k)} \}_{k \geq 1} \quad \quad \text{on } \quad
\cL^1 = \bigoplus_{k \geq 1} \cL^1_k
\end{equation}
with respect to the product of the weak* topologies of $\cL^1_k$.
Here \begin{equation}\label{eq:gammakt} \gamma_{t}^{(k)} (\bx_k;
\bx'_k) = \prod_{j=1}^k \ph_t (x_j) \overline{\ph_t (x'_j)}
\end{equation}
where $\ph_t$ is the solution of the non-linear Hartree
equation
\begin{equation}\label{eq:hartree2}
i\partial_t \ph_t = (1- \Delta)^{1/2} \ph_t + \lambda \left(
\frac{1}{|\, .\, |} * |\ph_t|^2 \right) \ph_t
\end{equation}
with initial data $\ph_{t=0} = \ph$.
\end{theorem}

\emph{Remarks.}
\begin{itemize}
\item[i)] The factorization assumption for the initial state
$\psi_{N}$ is only necessary in the limit $N \to \infty$. In other
words, if $\gamma^{(k)}_N$ denotes the $k$-particle marginal
associated with $\psi_N$, it is enough to assume that, for every
fixed $k \geq 1$,
\begin{equation*}
\tr \left| \gamma^{(k)}_{N} - \gamma^{(k)} \right| \to 0 \quad
\text{as  } N \to \infty
\end{equation*}
where $\gamma^{(k)} (\bx_k; \bx'_k) = \prod_{j=1}^k \ph (x_j)
\overline{\ph (x'_j)}$ for some $\ph \in H^1 (\bR^3)$. \item[ii)]
The result can also be easily generalized to the case where the
initial state is not in a pure state. If we assume the initial
density matrix to be given by
\[ \gamma_N (\bx;\bx') = \prod_{j=1}^N \gamma^{(1)} (x_j;x'_j),
\] for some $\gamma^{(1)} \in H^1 (\bR^3) \times H^1 (\bR^3)$, we can prove the
convergence (\ref{eq:conve}); in this case the limit $\Gamma_t =
\{ \gamma^{(k)}_t \}_{k\geq 1}$ is given by $\gamma^{(k)}_t
(\bx_k; \bx'_k) = \prod_{j=1}^k \gamma^{(1)}_t (x_j ; x'_j)$ where
$\gamma^{(1)}_t (x;x')$ is the solution of the nonlinear equation
\begin{equation*}
i\partial_t \gamma^{(1)} = \left[ (1-\Delta)^{1/2} + \lambda
\left( \frac{1}{|.|} * \rho_t \right) , \gamma^{(1)} \right] \quad
\text{with } \quad \rho_t (x) = \gamma^{(1)} (x;x)
\end{equation*}
with initial data $\gamma^{(1)}_{t=0} = \gamma^{(1)}$. This
equation is equivalent to the Hartree equation
(\ref{eq:hartree2}), if $\gamma^{(1)} (x;x') = \ph (x)
\overline{\ph}(x')$. \item[iii)] Although we state our theorem
specifically for the Coulomb potential $V(x) =\lambda |x|^{-1}$,
it is clear that our result and our proof still hold true for
$V(x) = A(x) |x|^{-1} + B (x)$, if, for example $A, B \in \cS
(\bR^3)$, the Schwarz class.
\end{itemize}

\bigskip

\begin{proof}
To simplify the notation we will assume $t \geq 0$. The proof for
$t < 0$ is similar.

\bigskip

\emph{Step 1. Introduction of the cutoffs.} For $\kappa
>0$ we define the regularized version of the initial one-particle
wave function $\ph \in H^1 (\bR^3)$ by
\begin{equation*}
\ph^{\kappa} = \exp(-\kappa |p|/N) \ph
\end{equation*}
with $p=-i\nabla$. Note that the regularized wave function
$\ph^{\kappa}$ depends on $N$. We use the notation
\begin{equation*}
\psi_{N}^{\kappa} (\bx) = \prod_{j=1}^N \ph^{\kappa} (x_j)
\end{equation*}
for the $N$ body wave function corresponding to the regularized
$\ph^{\kappa}$. We denote by $\psi_{N,t}^{\kappa}$ its time
evolution with respect to the physical Hamiltonian $H_N$
(\ref{eq:ham}), and by $\psi_{N,t}^{\kappa,\eps}$ its time
evolution with respect to the modified Hamiltonian $\wt H_{N}$
defined in (\ref{eq:cutoffham}). We also use the notation
$\gamma^{(k)}_{N,\kappa,t}$ and $\wt \gamma^{(k)}_{N,t} =
\gamma_{N,\kappa,\eps,t}^{(k)}$ for the family of marginal
distributions corresponding to $\psi_{N,t}^{\kappa}$ and
$\psi_{N,t}^{\kappa,\eps}$, respectively.

\bigskip

\emph{Step 2. A priori bounds.} In Section \ref{sec:apriori},
Theorem \ref{thm:apriori}, we prove that for any $k \geq 1$ and
for all $\eps , \kappa >0$ there exists $N_0 = N_0
(k,\eps,\kappa)$ such that
\begin{equation}\label{eq:aprio}
\tr \, |S_1 \dots S_k \, \wt \gamma^{(k)}_{N,t} S_k \dots S_1|
\leq C^k
\end{equation}
for all $N >N_0$ and for all $t \in \bR$. The constant $C$ is
independent of $k, \eps,\kappa, N,t$.

\bigskip

\emph{Step 3. Compactness of  $\wt \Gamma_{N,t}$.} Fix $T \geq 0$.
{F}rom Theorem \ref{thm:compact} it follows that, for every fixed
$k \geq 1$, the sequence $\wt \gamma^{(k)}_{N,t} \in C([0,T],
\cH_k)$ is compact with respect to the metric $\widehat \rho_k$
(defined in (\ref{eq:hatrho})). This implies that the sequence
$\wt \Gamma_{N,t} = \{ \wt \gamma^{(k)}_{N,t}\}_{k=1}^N$ is
compact in the space $C([0,T], \cH) = \bigoplus_{k \geq 1}
C([0,T], \cH_k)$ with respect to the product of the topologies
generated by the metrics $\widehat \rho_k$. Here we need the
following standard argument (choice of the diagonal subsequence)
to prove that, given a sequence $N_j$, with $N_j \to \infty$ for
$j \to \infty$, there exists a subsequence $M_j$ of $N_j$ with
$M_{j} \to \infty$ for $j \to \infty$, such that $\wt
\gamma^{(k)}_{M_{j},t}$ converges as $j \to\infty$, for all $k
\geq 1$; by the compactness of $\wt \gamma_{N,t}^{(1)}$ there is a
subsequence $\alpha_{1,j}$ of $N_j$, with $\alpha_{1,j} \to
\infty$ for $j \to \infty$, such that $\wt \gamma_{\alpha_{1,j}
,t}^{(1)}$ converges. Next, by the compactness of $\wt
\gamma_{N,t}^{(2)}$, there exists a subsequence $\alpha_{2,j}$ of
$\alpha_{1,j}$ with $\alpha_{2,j} \to \infty$ and such that $\wt
\gamma_{\alpha_{2,j} ,t}^{(2)}$ converges as $j \to \infty$
(clearly, since $\alpha_{2,j}$ is a subsequence of $\alpha_{1,j}$,
$\wt \gamma^{(1)}_{\alpha_{2,j},t}$ converges as well).
Inductively we can define subsequences $\alpha_{j,\ell}$ for all
$ell \geq 1$, and we can set $M_j = \alpha_{j,j}$: then $M_j \to
\infty$ for $j \to \infty$ and $\wt \gamma_{M_j,t}^{(k)}$
converges as $j \to \infty$, for all $k \geq 1$.

It also follows from Theorem \ref{thm:compact} that, if
$\Gamma_{\infty,t} = \{ \gamma_{\infty,t}^{(k)} \}_{k \geq 1} \in
C([0,T], \cH)$ denotes an arbitrary limit point of $\wt
\Gamma_{N,t}$, then we have
\begin{equation}\label{eq:aprioribou}
\tr \; |S_1 \dots S_k \gamma_{\infty,t}^{(k)} S_k \dots S_1| \leq
C^k
\end{equation}
for all $t \in [0,T]$ and for all $k \geq 1$ (with the same
constant $C$ as in (\ref{eq:aprio})).

\bigskip

\emph{Step 4. Convergence to the infinite hierarchy.} In Theorem
\ref{thm:convergence}, we prove that any limit point
$\Gamma_{\infty,t} \in C([0,T], \cH)$ of the sequence $\wt
\Gamma_{N,t}$ (with respect to the product of the topologies
$\wh\rho_k$) is a solution of the infinite hierarchy
(\ref{eq:BBGKYinf}), with initial value $\Gamma_{\infty,t=0} = \{
\gamma_{0}^{(k)} \}_{k\geq 1}$, where
\begin{equation}\label{eq:initcondi} \gamma^{(k)}_{0} (\bx_k ; \bx'_k) =
\prod_{j=1}^k \ph (x_j) \overline{\ph(x'_j)}\end{equation} (the
$\kappa$-dependence of $\ph^{\kappa}$ disappears in the limit $N
\to \infty$).

\bigskip

\emph{Step 5. Uniqueness of the infinite hierarchy.} In Theorem
\ref{thm:unique} we demonstrate that for any given $\Gamma_{0} =
\{ \gamma^{(k)}_{0} \}_{k \geq 1} \in \cH$, satisfying the
estimate
\begin{equation*}
\| \gamma^{(k)}_0 \|_{\cH_k} = \tr \; |S_1 \dots S_k
\gamma^{(k)}_0 S_k \dots S_1| \leq C^k
\end{equation*}
for some $C > 0$, there is at most one solution $\Gamma_{\infty,t}
= \{ \gamma_{\infty,t}^{(k)} \}_{k \geq 1} \in C([0,T], \cH)$ of
the infinite hierarchy (\ref{eq:BBGKYinf}) which satisfies
$\Gamma_{\infty,t=0} = \Gamma_0$ and
\begin{equation}\label{eq:bou}
\| \gamma^{(k)}_{\infty,t} \|_{\cH_k} \leq C^k
\end{equation}
for all $t \in [0,T]$ and $k \geq 1$. This, together with
(\ref{eq:aprioribou}), and with the fact that, by Step 4, every
limit point of $\wt \Gamma_{N,t}$ satisfies the infinite hierarchy
(\ref{eq:BBGKYinf}), immediately implies that the sequence $\wt
\Gamma_{N,t}$ converges for $N \to \infty$ (a compact sequence
with only one limit point is always convergent). Next, we note
that the family of density $\Gamma_t = \{ \gamma_t^{(k)}\}_{k \geq
1}$, with $\gamma_t^{(k)}$ defined by (\ref{eq:gammakt}),
satisfies (\ref{eq:BBGKYinf}) and the initial condition
(\ref{eq:initcondi}) (the existence and uniqueness of a global
solution of the nonlinear Hartree equation (\ref{eq:hartree2}),
for $\lambda > \lambda_{\text{crit}}$, is proven in \cite{L}).
Moreover, it also satisfies the bound (\ref{eq:bou}). In fact, by
Lemma \ref{lm:hartree},
\[ \| \gamma_t^{(k)} \|_{\cH_k} = \tr \;
|S_1 \dots S_k \gamma_t^{(k)} S_k \dots S_1| = (\ph_t,
(1-\Delta)^{1/2} \ph_t)^k \leq C^k \, . \] It follows that
\begin{equation*}
\wt \Gamma_{N,t} \to \Gamma_t \quad \quad \text{as } N \to \infty
\end{equation*}
on $C([0,T], \cH)$, with respect to the product of the topologies
induced by $\wh\rho_k$, for $k \geq 1$. This, in particular,
implies that, for every fixed $t\in [0,T]$, we have $\wt
\Gamma_{N,t} \to \Gamma_t$ in $\cH$, with respect to the product
of the weak* topologies of $\cH_k$ (which, since the sequence $\wt
\gamma^{(k)}_{N,t}$ is uniformly bounded in $N$, are equivalent to
the topologies generated by the metrics $\rho_k$). In turn this
implies that, for every fixed $t \in [0,T]$, we have
\begin{equation}\label{eq:step5}
\wt \Gamma_{N,t} \to \Gamma_t \quad \quad \text{on } \quad \cL^1 =
\bigoplus_{k \geq 1} \cL^1_k,
\end{equation} with respect to the product of the weak* topologies
on $\cL^1_k$, for $k \geq 1$.

\bigskip

\emph{Step 6. Removal of the cutoffs.} Recall that $\psi_{N,t}$
denotes the evolution of the initial function $\psi_N$ generated
by the Hamiltonian $H_N$, defined in (\ref{eq:ham}). Moreover,
$\psi_{N,t}^{\kappa}$ and $\psi_{N,t}^{\kappa,\eps}$ denote the
time evolution of the regularized initial wave function
$\psi_N^{\kappa}$ with respect to the dynamics generated by the
original Hamiltonian $H_N$ and, respectively, by the regularized
Hamiltonian $\wt H_{N}$ (defined in (\ref{eq:cutoffham})). We
remind that $\gamma_{N,t}^{(k)}$, $\gamma_{N,\kappa,t}^{(k)}$ and
$\wt \gamma_{N,t}^{(k)} = \gamma_{N,\kappa,\eps,t}^{(k)}$ denote
the marginal distributions associated with $\psi_{N,t}$, to
$\psi_{N,t}^{\kappa}$ and, respectively, with
$\psi_{N,t}^{\kappa,\eps}$. In Proposition \ref{prop:comp}, we
show that
\begin{equation*}
\| \psi_{N,t}^{\kappa} - \psi_{N,t}^{\eps,\kappa} \| \leq C t \,
\eps^{1/4}
\end{equation*}
for every $N$ large enough (depending on $\eps,\kappa$). This
clearly implies that \begin{equation}\label{eq:rem1} \tr \;
|\gamma^{(k)}_{N,t,\kappa} - \gamma^{(k)}_{N,t,\kappa,\eps}| \leq
C t \, \eps^{1/4}\, \end{equation} for all $k \geq 1$ and for all
$N$ large enough (depending on $\eps, \kappa$). Moreover, in
Proposition \ref{prop:kappa}, we show that
\begin{equation*}
\| \psi_{N,t}^{\kappa} - \psi_{N,t}\| = \| \psi_{N}^{\kappa} -
\psi_N \| \leq C \kappa
\end{equation*}
for all $N$ and $t\in \bR$. Therefore
\begin{equation}\label{eq:rem2}
\tr \; |\gamma^{(k)}_{N,\kappa,t} - \gamma^{(k)}_{N,t}| \leq C
\kappa
\end{equation}
for all $k\geq 1$, $t\in \bR$ and $N\geq 1$.

\bigskip

\emph{Step 7. Conclusion of the proof.} For every fixed $k \geq 1$
and $t\in [0,T]$, and for every compact operator $J^{(k)} \in
\cK_k$, we have
\begin{equation*}\begin{split}
\left| \tr \; J^{(k)} \, \left(\gamma_{N,t}^{(k)} -
\gamma_{t}^{(k)} \right) \right| \leq &\;  \| J^{(k)} \| \, \tr \;
\left|\gamma_{N,t}^{(k)} - \gamma_{N,\kappa,t}^{(k)}\right| + \|
J^{(k)} \| \, \tr
\;  \left|\gamma_{N,\kappa,t}^{(k)} - \gamma_{N,\kappa,\eps,t}^{(k)}\right| \\
&+ \left| \tr \; J^{(k)} \left( \gamma^{(k)}_{N,\kappa,\eps,t} -
\gamma_t^{(k)} \right) \right|,
\end{split}
\end{equation*}
where $\gamma_t^{(k)}$ is defined in (\ref{eq:gammakt}). By
(\ref{eq:rem1}) and (\ref{eq:rem2}), for any fixed $\delta
>0$, we can find $\kappa$ and $\eps$ small enough such that both
the first and the second term on the r.h.s. of the last equation
are smaller than $\delta /3$, uniformly in $N$, if $N$ is large
enough. Finally, by (\ref{eq:step5}), the last term can be made
smaller than $\delta/3$ by choosing $N$ sufficiently large (and
keeping $\eps$ and $\kappa$ fixed). Thus, for every fixed $t \in
[0,T]$, $k\geq 1$, $J^{(k)} \in \cK_k$, and $\delta >0$, we have
\begin{equation*} \left| \tr \; J^{(k)} \left(\gamma_{N,t}^{(k)} -
\gamma^{(k)}_{\infty,t}\right) \right| \leq \delta \end{equation*}
for all $N$ large enough. In other words, $\Gamma_{N,t} \to
\Gamma_{\infty,t}$ with respect to the product of the weak*
topologies of $\cL^1_k$, $k \geq 1$, for every fixed $t \in
[0,T]$. Since $T< \infty$ is arbitrary, this completes the proof
of the theorem.
\end{proof}

\section{A--Priori Estimates}
\label{sec:apriori} \setcounter{equation}{0}

The aim of this section is to prove an priori bound, of the form
\begin{equation}\label{eq:ap}
\tr \; |S_1 \dots S_k \wt \gamma^{(k)}_{N,t} S_k \dots S_1| \leq
C^k \, , \end{equation} for the marginals $\wt \gamma^{(k)}_{N,t}$
associated with the solution $\psi_{N,t}^{\eps,\kappa}$ of the
modified Schr\"odinger equation (\ref{eq:schrcut}) with
regularized initial data $\psi_{N,t=0}^{\eps,\kappa} =
\psi_N^{\kappa}$ (recall that $S_j = (1-\Delta_j)^{1/4}$). To
prove (\ref{eq:ap}) we derive an upper and a lower bound for
powers of $\wt H_N$ in terms of corresponding powers of the
kinetic energy.

\begin{proposition}[Lower Bound for $\wt H_N^k$]\label{prop:lower}
We distinguish two cases.
\begin{itemize}
\item[i)] If $\lambda >0$: For every $k \geq 0$, and $C<1$, there
exists $N_0 = N_0 (k,C,\eps)$ (depending also on the cutoff $\eps$
in the potential) such that
\begin{equation}\label{eq:cl1}
(\psi , \wt H_N^k \psi ) \geq C^k \left\{  N^k (\psi, S_1^2 \dots
S_k^2 \psi) + N^{k-1} (\psi, S_1^4 S_2^2 \dots S_{k-1}^2 \psi)
\right\}
\end{equation}
for all $N>N_0$, and for all $\psi$ invariant with respect to
permutations. \item[ii)] If $-4/\pi < \lambda <0$: For every $k
\geq 0$, and $C<1$, there exists $N_0 = N_0 (k,C,\eps)$ such that
\begin{equation}\label{eq:cl2}
\begin{split}
(\psi , \wt H_N^k \psi ) &\geq C^k (1 + \frac{\pi}{4}
\lambda)^{[k/2]} \, \left\{ N^k (\psi, S_1^2 \dots S_k^2 \psi) +
N^{k-1} (\psi, S_1^4 S_2^2 \dots S_{k-1}^2 \psi) \right\}
\end{split}
\end{equation} for all $N>N_0$, and for all $\psi$ symmetric
with respect to permutations. Here $[k/2] = k/2$ if $k$ is even,
and $[k/2] = (k+1) /2$ if $k$ is odd.
\end{itemize}
\end{proposition}

\begin{proof}
We use a two-step induction over $k$. For $k=0$ the statement is
trivial. For $k=1$, and $\lambda>0$, the claim follows by the
positivity of the potential. For $k=1$ and $\lambda <0$, it
follows from the operator inequality \[ \frac{1}{|x_1 -x_2|} \leq
\frac{\pi}{2} S_1^2
\] (see Lemma \ref{lm:norm-bound}, part (i))
and by the symmetry of $\psi$ with respect to permutations. Next
we assume that the claim holds true for $k= n$, and we prove it
for $k=n+2$. We consider the case $\lambda <0$ (if $\lambda
>0$ the proof is easier), and we assume that $n$ is even (for odd $n$ the proof is
analogous). By the induction assumption, for any $C <1$, there
exists $N_0 (n,C,\eps)$ such that
\begin{equation}\label{eq:indassum}
(\psi , \wt H_N^{n+2} \psi) = (\psi, \wt H_N \wt H_N^n \wt H_N
\psi ) \geq C^n N^n (1+\frac{\pi}{4}\lambda)^{n/2} \, (\psi, \wt
H_N S_1^2 \dots S_n^2 \wt H_N \psi) \,
\end{equation}
for all $N > N_0 (n,C,\eps)$, and for all $\psi$ symmetric with
respect to permutations (note that also $\wt H_N \psi$ is
symmetric). Here we are neglecting, because of its positivity, the
contribution arising from the second term in the parenthesis on
the r.h.s. of (\ref{eq:cl2}). Writing
\[ \wt H_N = \sum_{j=n+1}^N S_j^2 + h_N \quad \quad \text{with}
\quad \quad h_N = \sum_{j=1}^n S_j^2 + \frac{\lambda}{N}
\sum_{i<j}^N \frac{1}{|x_i - x_j| +\eps N^{-1}}, \] equation
(\ref{eq:indassum}) implies that
\begin{equation*}
\begin{split}
(\psi , \wt H_N^{n+2} \psi) \geq \; & C^n N^n
(1+\frac{\pi}{4}\lambda)^{n/2}  \sum_{i,j \geq
n+1} (\psi, S_i^2 S_1^2 \dots S_n^2 S_j^2 \psi)   \\
&+ C^n N^n (1+\frac{\pi}{4}\lambda)^{n/2} \sum_{i \geq n+1}
\left(( \psi, S_i^2 S_1^2 \dots S_n^2 h_N \psi) +
\text{c.c.}\right) \,,
\end{split}
\end{equation*}
where $c.c.$ denotes the complex conjugate. Using the permutation
symmetry we find
\begin{equation}\label{eq:enest1}
\begin{split}
(\psi , \wt H_N^{n+2} \psi) \geq \; C^n &N^n (1 + \frac{\pi}{4}
\lambda)^{n/2}\\ \times & \Big[(N-n)(N-n-1) (\psi, S_1^2 \dots
S_{n+2}^2 \psi)
+ (N-n) (\psi, S_1^4 S_2^2 \dots S_{n+1}^2 \psi)  \\
&+  \lambda \frac{(N-n)(n+1)n}{2N} (\psi, V^{\eps}_{1,2} S_1^2
\dots S_{n+1}^2 \psi ) + \text{c.c.} \\ &+ \lambda
\frac{(N-n)(n+1) (N-n-1)}{N}
(\psi, V^{\eps}_{1,n+2} S_1^2 \dots S_{n+1}^2 \psi ) + \text{c.c.}  \\
& + \lambda \frac{(N-n)(N-n-1)(N-n-2)}{2N} (\psi,
V^{\eps}_{n+2,n+3} S_1^2 \dots S_{n+1}^2 \psi) + \text{c.c.} \Big]
\end{split}
\end{equation}
where we use the notation $V_{i,j}^{\eps} = 1/ (|x_i -x_j| + \eps
N^{-1})$. We first estimate the last term (for positive $\lambda$,
this term is positive, and thus can be neglected). Since
$V_{n+2,n+3}^{\eps}$ commutes with $S_j$, for all $j \leq n+1$, we
have
\begin{equation}\label{eq:Vn+2n+3}
(\psi, V^{\eps}_{n+2,n+3} S_1^2 \dots S_{n+1}^2 \psi) = (\psi ,
S_1 \dots S_{n+1} V^{\eps}_{n+2,n+3} S_{n+1} \dots S_1 \psi) \leq
\frac{\pi}{2} \, (\psi , S_1^2 \dots S_{n+2}^2 \psi)
\end{equation}
because $V^{\eps}_{n+2,n+3} \leq |x_{n+2} - x_{n+3}|^{-1} \leq
(\pi/2) \, S_{n+2}^2$ (Lemma \ref{lm:norm-bound}, part (i)). Now
we turn our attention to the fourth term in (\ref{eq:enest1}). We
have
\begin{equation}\label{eq:V1n+2}
\begin{split}
\left|(\psi , V^{\eps}_{1,n+2} S_1^2 \dots S_{n+1}^2 \psi) \right|
= \; &\left|(\psi , S_{n+1}  \dots  S_{2}
V^{\eps}_{1,n+2} S_1^2 S_2 \dots S_{n+1} \psi) \right| \\
= \; & \left|(\psi, S_{n+2} S_{n+1} \dots S_{1} \{ S_1^{-1}
S_{n+2}^{-1} V^{\eps}_{1,n+2} S_1 S_{n+2}^{-1} \} S_1 \dots
S_{n+1} S_{n+2} \psi)\right| \\ \leq \; & \const \; (\psi , S_1^2
\dots S_{n+2}^2 \psi)
\end{split}
\end{equation}
where we used that, by Lemma \ref{lm:norm-bound}, part (ii), the
norm $\| S_1^{-1} S_{n+2}^{-1} V^{\eps}_{1, n+2} S_1 S_{n+2}^{-1}
\|$ is finite, uniformly in $\eps$. Finally we consider the third
term in (\ref{eq:enest1}). To this end we note that
\begin{equation*}
\begin{split}
(\psi , V^{\eps}_{1,2} S_1^2 \dots S_{n+1}^2 \psi) = \; &(\psi ,
S_{n+1} \dots  S_{3} V^{\eps}_{1,2} S_1^2 S_2^2 S_3 \dots S_{n+1}
\psi) \\ = \; & (\psi, S_{n+1} \dots S_3 S_2 S_1^2 \{ S_1^{-2}
S_2^{-1} V^{\eps}_{1,2} S_2 \} S^2_1 S_2 \dots S_{n+1} \psi) \, .
\end{split}
\end{equation*}
For every $\delta >0$ there exists, by Lemma \ref{lm:norm-bound},
part (iii), a constant $C_{\eps,\delta} >0$ (depending also on
$\eps$) such that
\begin{equation}\label{eq:V12}
\left|(\psi , V^{\eps}_{1,2} S_1^2 \dots S_{n+1}^2 \psi)\right|
\leq C_{\eps , \delta} N^{\delta} (\psi, S_1^4 S_2^2 \dots
S_{n+1}^2 \psi) \, .
\end{equation}
We fix $0< \delta < 1$. Then, inserting (\ref{eq:Vn+2n+3}),
(\ref{eq:V1n+2}), and (\ref{eq:V12}) into (\ref{eq:enest1}), we
find
\begin{equation*}
\begin{split}
(\psi , \wt H_N^{n+2} \psi) \geq \; &C^n N^n (1 + \frac{\pi}{4}
\lambda)^{n/2} \left\{ (N-n)(N-n-1) + \frac{\pi}{4}
\lambda \frac{(N-n)(N-n-1)(N-n-2)}{N} \right.\\
&\hspace{3cm} \left. - \,\const \, \frac{(N-n)(n+1) (N-n-1)}{N}
\right\} (\psi , S_1^2 \dots S_{n+2}^2 \psi) \\ &+ C^n N^n (1 +
\frac{\pi}{4}\lambda)^{n/2} \left\{ (N-n) - C_{\eps, \delta}
N^{\delta} \right\} (\psi, S_1^4 S_2^2 \dots
S_{n+1}^2 \psi) \\
= \; &C^n N^{n+2} (1+ \frac{\pi}{4} \lambda)^{n/2}
\frac{(N-n)(N-n-1)}{N^2} \\ &\hspace{3cm} \times \left( 1
+\frac{\pi}{4}\lambda \frac{N-n-2}{N} -
\const \, \frac{n+1}{N}\right) (\psi, S_1^2 \dots S_{n+2}^2 \psi) \\
&+ C^n N^{n+1} (1+ \frac{\pi}{4} \lambda)^{n/2} \frac{N-n}{N} \,
\left(1- C_{\eps, \delta} \frac{N^{\delta}}{N-n} \right) (\psi,
S_1^4 S_2^2 \dots S_{n+1}^2 \psi) \, .
\end{split}
\end{equation*}
Next, since $C<1$, we can find $N_0 (n+2,C, \eps) > N_0
(n,C,\eps)$ such that the four inequalities
\begin{equation*}
\frac{(N-n)(N-n-1)}{N^2} \geq C, \quad \left( 1
+\frac{\pi}{4}\lambda \frac{N-n-2}{N} - \const \,
\frac{n+1}{N}\right) \geq  C \left(1+\frac{\pi}{4}\lambda\right)
\quad \text{and} \end{equation*}
\begin{equation*}
\frac{N-n}{N} \geq C,  \quad \quad \left(1- C_{\eps, \delta}
\frac{N^{\delta}}{N-n} \right) \geq C
\end{equation*}
are satisfied for all $N > N_0 (n+2 ,C,\eps)$. This implies that
\begin{equation*}
(\psi , \wt H_N^{n+2} \psi) \geq C^{n+2}
\left(1+\frac{\pi}{4}\lambda\right)^{(n+2)/2} \left\{ N^{n+2}
(\psi , S_1^2 \dots S_{n+2}^2 \psi) + N^{n+1} (\psi, S_1^4 S_2^2
\dots S_{n+1}^2 \psi) \right\}
\end{equation*}
for all $N > N_0 (n+2,C)$, and completes the proof of the
proposition.
\end{proof}

In the next proposition, we prove an upper bound for powers of
$\wt H_N$.

\begin{proposition}[Upper Bound for $\wt H_N^k$]
\label{prop:upper} For $\ell \geq 1$, we use the notation
$\balpha_{\ell} = (\alpha_1, \dots , \alpha_{\ell}) \in
\bN^{\ell}_+$, and $|\balpha_{\ell}| = \alpha_1 + \dots +
\alpha_{\ell}$. Assume $C$ is large enough (depending only on
$\lambda$). Then, for every fixed $k \geq 0$, $\eps
>0$ and for every $a
>0$ there exists $0 \leq C(k,\eps,a) < \infty$ (also depending on $\lambda$)
such that
\begin{equation}\label{eq:claim}
(\psi , \wt H_N^k \psi ) \leq C^k (\psi , \left( S_1^2 + \dots
S_N^2 \right)^k \psi) + C(k,\eps,a) \, N^{a} \sum_{\ell=1}^{k-1}
\sum_{\balpha_{\ell} : |\balpha_{\ell}|=k} N^{\ell} \, (\psi,
S_1^{2\alpha_1} \dots S_{\ell}^{2 \alpha_{\ell}} \psi)
\end{equation}
for all $N$ and for all $\psi$ symmetric with respect to
permutations of the $N$ particles.
\end{proposition}

\begin{proof}
We prove the proposition by a two step induction over $k$. The
statement is clear for $k=0$ and $k=1$ (in this case one can
choose $C(k=1,\eps,a)=0$). We assume now that the statement holds
for $k=n$, and we prove it for $k=n+2$. {F}rom the induction
assumption, with $a$ replaced by $a(n-1)/n$, and since $\psi$ and
$\wt H_N \psi$ are symmetric with respect to permutations, we have
\begin{equation}\label{eq:up1}
\begin{split}
(\psi , \wt H_N^{n+2} \psi) =\; & (\psi , \wt H_N \wt H_N^{n} \wt
H_N \psi) \\ \leq \; &C^n (\psi ,\wt H_N (S_1^2 + \dots +S_N^2)^n
\wt H_N \psi) \\ &+ C(n,\eps,a(n-1)/n) \, N^{a(n-1)/n}
\sum_{r=1}^{n-1} \sum_{\balpha_{r} : |\balpha_{r}|=k} N^{r}\,
(\psi, \wt H_N \, S_1^{2\alpha_1} \dots S_{r}^{2 \alpha_{r}} \wt
H_N \psi)\,.
\end{split}
\end{equation}
We start by considering the first term on the r.h.s. of the last
equation. Applying a Schwarz inequality and using again the
permutation symmetry, we find
\begin{equation*}
\begin{split}
(\psi ,\wt H_N (S_1^2 &+ \dots +S_N^2)^n \wt H_N \psi) \\ &\leq \;
2 (\psi, (S_1^2 + \dots + S_N^2)^{n+2} \psi) + N (N-1) \lambda^2
(\psi, V_{1,2}^{\eps} ( S_1^2 + \dots + S_N^2)^n V_{1,2}^{\eps}
\psi)\,.
\end{split}
\end{equation*}
The weighted Minkowski inequality \[ (S_1^2 + \dots + S_N^2)^n
\leq 2 (S_3^2 + \dots + S_N^2)^n + c(n) (S_1^{2n} + S_2^{2n}) \]
(where the constant $c(n)$ can be chosen as $c^n n^n$, for some
$c>0$) implies that
\begin{equation}\label{eq:up2}
\begin{split}
(\psi ,\wt H_N (S_1^2 + \dots +S_N^2)^n \wt H_N \psi) \leq \; &2\,
(\psi, (S_1^2 + \dots + S_N^2)^{n+2} \psi) \\ &+ 2N (N-1)
\lambda^2 (\psi, V_{1,2}^{\eps} ( S_3^2 + \dots + S_N^2)^n
V_{1,2}^{\eps} \psi) \\ &+ 4 N^2 \lambda^2 c(n) (\psi,
V_{1,2}^{\eps} S_1^{2n} V_{1,2}^{\eps} \psi)\, .
\end{split}
\end{equation}
The second term on the r.h.s. of
the last equation can be bounded by
\begin{equation}\label{eq:up2a}
\begin{split}
2N (N-1) \lambda^2 (\psi, V_{1,2}^{\eps} ( S_3^2 + \dots +
S_N^2)^n V_{1,2}^{\eps} \psi) \leq \; & C \, N (N-1) (\psi, (S_3^2
+ \dots S_N^2)^n S_1^2 S_2^2 \psi) \\ \leq \; &C \, (\psi, (S_1^2
+ \dots + S_N^2)^{n+2} \psi)
\end{split}
\end{equation}
because $V_{1,2}^{\eps}$ commutes with $(S_3^2 + \dots +S_N^{2})$,
and $(V_{1,2}^{\eps})^2 \leq |x_1 - x_2|^{-2} \leq C \, S_1^2
S_2^2$ (see Lemma \ref{lm:norm-bound}) and because $N(N-1) (\psi ,
S_1^2 S_2^2 \psi) \leq (\psi, (S_1^2 + \dots +S_N^2)^2 \psi)$ for
permutation invariant $\psi$. Next we consider the last term on
the r.h.s. of (\ref{eq:up2}). Using Lemma \ref{lm:comm}, for every
$a>0$, $n \geq 0$ and $\eps>0$, we find a constant $D(n,\eps,a)$
such that
\begin{equation*}
\begin{split}
N^2 \, V_{1,2}^{\eps} S_1^{2n} V_{1,2}^{\eps} \leq D(n,\eps,a)
N^{a} \, N^2 \sum_{m=1}^n S_1^{2(n-m)} S_1^4 S_2^2 N^{m-1} \, .
\end{split}
\end{equation*}
In the sum over $m$, we can add more factors of $S^2_j$, so that
the order of each monomial is $2(n+2)$ (and we can compare this
contribution with the sum in the r.h.s. of (\ref{eq:claim})). More
precisely we can bound
\[ S_1^{2(n-m)} S_1^4 S_2^2 = S_1^{2(n-m +2)} S_2^2 \leq
S_1^{2(n-m+2)} S_2^2 S_3^2 \dots S_m^2 S_{m+1}^2
\] because $S_j^2 \geq 1$. Hence, for every $a>0$,
there exists a constant $C_1 (n, \eps,a)$ (also depending on
$\lambda$) such that
\begin{equation*}
\begin{split}
4c(n)\lambda^2 N^2 (\psi, V_{1,2}^{\eps} &S_1^{2n} V_{1,2}^{\eps}
\psi) \leq C_1(n,\eps,a) N^{a} \sum_{\ell=1}^{n+1}
\sum_{\balpha_{\ell}: |\balpha_{\ell}| = n+2}  N^{\ell} \, (\psi,
S_1^{2\alpha_1} \dots S_{\ell}^{2\alpha_{\ell}} \psi) \, .
\end{split}
\end{equation*}
{F}rom the last equation, together with (\ref{eq:up2}),
(\ref{eq:up2a}), it follows that
\begin{equation}\label{eq:up2b}
\begin{split}
(\psi,\wt H_N (S_1^2 + \dots + S_N^2)^n \wt H_N \psi) \leq \; &C^2
(\psi ,(S_1^2 + \dots + S_N^2)^{n+2} \psi) \\ &+ C_1(n,\eps,a) \,
N^a  \sum_{\ell=1}^{n+1} \sum_{\balpha_{\ell}: |\balpha_{\ell}| =
n+2} (\psi, S_1^{2\alpha_1} \dots S_{\ell}^{2\alpha_{\ell}} \psi)
N^{\ell}
\end{split}
\end{equation}
provided $C$ is large enough (depending only on $\lambda$).

Next we consider the second term on the r.h.s. of (\ref{eq:up1}).
Applying the Schwarz inequality, we find
\begin{equation}\label{eq:up3}
\begin{split}
\sum_{r=1}^{n-1} \sum_{\balpha_{r} : |\balpha_{r}|=n} N^{r}\,
(\psi, \wt H_N \, S_1^{2\alpha_1} \dots &S_{r}^{2 \alpha_{r}} \wt
H_N \psi) \\ \leq \; &2 \sum_{r=1}^{n-1} \sum_{\balpha_{r} :
|\balpha_{r}|=n} N^{r}\, (\psi, (S_1^2 + \dots + S_N^2)^2
S_1^{2\alpha_1} \dots S_{r}^{2\alpha_{r}} \psi) \\ &+
\sum_{r=1}^{n-1} \sum_{\balpha_{r} : |\balpha_{r}|=n} N^{r}\,
\sum_{i<j} (\psi, V_{i,j}^{\eps}S_1^{2\alpha_1} \dots S_{r}^{2
\alpha_{r}} V_{i,j}^{\eps} \psi)\,.
\end{split}
\end{equation}
Using the permutation symmetry, the first term on the r.h.s. of
last equation can be rewritten as
\begin{equation}\label{eq:up3b}
\begin{split}
\sum_{r=1}^{n-1} &\sum_{\balpha_{r} : |\balpha_{r}|=n} N^{r}\,
(\psi, (S_1^2 + \dots + S_N^2)^2 S_1^{2\alpha_1} \dots
S_{r}^{2\alpha_{r}} \psi) \\ = \; &\sum_{r=1}^{n-1}
\sum_{\balpha_{r} : |\balpha_{r}| = n} N^{r} \left\{ r \, (\psi,
S_1^{2(\alpha_1+2)} S_2^{2\alpha_2} \dots S_{r}^{2\alpha_{r}}
\psi) \right. + 2 r (r-1)\, (\psi, S_1^{2(\alpha_1+1)}
S_2^{2(\alpha_2+1)} S_3^{2\alpha_3} \dots S_{r}^{2\alpha_{r}}
\psi) \\ &\hspace{3cm} + 2 r (N-r) (\psi, S_1^{2(\alpha_1 + 1)}
\dots S_{r}^{2\alpha_{r}} S_{r+1}^2 \psi) + (N-r) (\psi,
S_1^{2\alpha_1 } \dots S_{r}^{2\alpha_{r}} S_{r+1}^4 \psi)
\\ & \hspace{3cm}\left. + 2 (N-r) (N-r-1) (\psi,
S_1^{2\alpha_1} \dots S_{r}^{2\alpha_{r}} S_{r+1}^2 S_{r+2}^2
\psi) \right\}
\end{split}
\end{equation}
which clearly implies that
\begin{equation*} \begin{split}
\sum_{r=1}^{n-1} &\sum_{\balpha_{r} : |\balpha_{r}|=n} N^{r}\,
(\psi, (S_1^2 + \dots + S_N^2)^2 S_1^{2\alpha_1} \dots
S_{r}^{2\alpha_{r}} \psi) \leq \;  \sum_{\ell=1}^{n+1}
\sum_{\balpha_{\ell} : |\balpha_{\ell}|=n+2} N^{\ell} (\psi,
S_1^{2\alpha_1} \dots S_{\ell}^{2\alpha_{\ell}} \psi)\,.
\end{split}
\end{equation*}
The second term on the r.h.s. of (\ref{eq:up3}), on the other
hand, can be rewritten as
\begin{equation}\label{eq:up4}
\begin{split}
\sum_{r=1}^{n-1} &\sum_{\balpha_{r} : |\balpha_{r}|=n} N^{r}\,
\sum_{i<j} (\psi, V_{i,j}^{\eps}S_1^{2\alpha_1} \dots S_{r}^{2
\alpha_{r}} V_{i,j}^{\eps} \psi) \\  \; &\hspace{1cm}
=\sum_{r=1}^{n-1} \sum_{\balpha_{r} : |\balpha_{r}|=n} N^{r}
\left\{ \frac{r(r-1)}{2} (\psi, V_{1,2}^{\eps} S_1^{2\alpha_1}
\dots S_{r}^{2\alpha_{r}} V_{1,2}^{\eps} \psi) \right. \\
&\hspace{3cm}+ r (N-r) (\psi, V_{1,r+1}^{\eps} \, S_1^{2\alpha_1}
\dots S_{r}^{2\alpha_{r}} V_{1,r+1}^{\eps} \psi)
\\ &\hspace{3cm} \left. + \frac{1}{2} (N-r)(N-r-1) (\psi,
V_{r+1,r+2}^{\eps} \, S_1^{2\alpha_1} \dots S_{r}^{2\alpha_{r}}
V_{r+1,r+2}^{\eps} \psi) \right\}\,.
\end{split}
\end{equation}
To bound the last contribution we use that, since
$V_{r+1,r+2}^{\eps}$ commutes with $S_1, \dots , S_{r}$, and since
$V_{r+1,r+2}^{\eps} \leq C \, S_{r+1}^2 S_{r+2}^2$,
\[ (\psi, V_{r+1,r+2}^{\eps} \,
S_1^{2\alpha_1} \dots S_{r}^{2\alpha_{r}} V_{r+1,r+2}^{\eps} \psi)
\leq C \, (\psi, S_1^{2\alpha_1} \dots S_{r}^{2\alpha_{r}}
S_{r+1}^{2} S_{r+2}^2 \psi). \] Hence
\begin{equation}\label{eq:up4b}
\begin{split}
\sum_{r=1}^{n-1} \sum_{\balpha_{r} : |\balpha_{r}|=n} N^{r} (N-r)
(N-r-1) &(\psi, V_{r+1,r+2}^{\eps} S_1^{2\alpha_1} \dots
S_{r}^{2\alpha_{r}} V_{r+1,r+2}^{\eps} \psi) \\ &\leq C
\sum_{\ell=1}^{n+1} \sum_{\balpha_{\ell} : |\balpha_{\ell}|=n+2}
N^{\ell} (\psi, S_1^{2\alpha_1} \dots S_{\ell}^{2\alpha_{\ell}}
\psi) \, .
\end{split}
\end{equation}
As for the second term in (\ref{eq:up4}), we write
\[ (\psi, V_{1,r+1}^{\eps} \, S_1^{2\alpha_1} \dots
S_{r}^{2\alpha_{r}} V_{1,r+1}^{\eps} \psi) = (\psi, S_2^{\alpha_2}
\dots S_{r}^{\alpha_{r}} V_{1,r+1}^{\eps} S_1^{2\alpha_1}
V_{1,r+1}^{\eps} S_{r}^{\alpha_{r}} \dots S_2^{\alpha_2} \psi)\,\]
and then we observe that, by Lemma \ref{lm:comm}, for every $a>0$
we can find $D(n,\eps,a)$ such that
\begin{equation}\label{eq:up5}
\begin{split}
V_{1,r+1}^{\eps} S_1^{2\alpha_1} V_{1,r+1} \leq D(n,\eps,a)
N^{a/n} \sum_{m=1}^{\alpha_1} S_1^{2(\alpha_1 - m + 2)} S_{r+1}^2
\, N^{m-1}\, .
\end{split}
\end{equation}
This implies that, for a suitable constant $C_2(n,\eps,a)$,
\begin{equation}\label{eq:up6}
\begin{split}
\sum_{r=1}^{n-1} &\sum_{\balpha_{r} : |\balpha_{r}|=n} N^{r} r
(N-r) (\psi, V_{1,r+1}^{\eps} \, S_1^{2\alpha_1} \dots
S_{r}^{2\alpha_{r}} V_{1,r+1}^{\eps} \psi) \\ &\leq C_2 (n,\eps,a)
\, N^{a/n} \sum_{r=1}^{n-1} \sum_{\balpha_{r} : |\balpha_{r}|=n}
\sum_{m=1}^{\alpha_1} N^{r+ m} (\psi, S_1^{2(\alpha_1-m+2)}
S_2^{2\alpha_2} \dots S_{r}^{2\alpha_{r}} S_{r+1}^2 \psi)
\\ &\leq C_2 (n,\eps,a) N^{a/n} \sum_{\ell=1}^{n+1}
\sum_{\balpha_{\ell} : |\balpha_{\ell}|=n+2} N^{\ell} (\psi,
S_1^{2\alpha_1} \dots S_{\ell}^{2\alpha_{\ell}} \psi) \, .
\end{split}
\end{equation}
In the last inequality we used that in the sum over $m$ we can add
more derivatives (until the total order of each summand is
$2(n+2)$), according to the bound
\[ N^{r+m} \, S_1^{2(\alpha_1-m+2)} S_2^{2\alpha_2} \dots
S_{r}^{2\alpha_{r}} S_{r+1}^2 \leq N^{r+m} \,
S_1^{2(\alpha_1-m+2)} S_2^{2\alpha_2} \dots S_{r}^{2\alpha_{r}}
S_{r+1}^2 S_{r+2}^2 \dots S_{r+m}^2 .\] Moreover, note that $r+m
\leq r + \alpha_1 = r + n - (\alpha_2 + \dots + \alpha_{r}) \leq
r+ n - (r-1) \leq n+1$ (this explains why the sum over the index
$\ell$, in the last line of (\ref{eq:up6}), only goes up to
$n+1$).

Finally, we consider the first term in (\ref{eq:up4}). We have,
using the symmetry with respect to permutations,
\begin{equation*}
\begin{split}
(\psi, V_{1,2}^{\eps} \, S_1^{2\alpha_1} \dots S_{r}^{2\alpha_{r}}
V_{1,2}^{\eps} \psi) &= (\psi, S_3^{\alpha_3} \dots
S_{r}^{\alpha_{r}} V_{1,2}^{\eps} S_1^{2\alpha_1} S_2^{2\alpha_2}
V_{1,2}^{\eps} S_{r}^{\alpha_{r}} \dots S_3^{\alpha_2} \psi)
\\ &\leq (\psi, S_3^{\alpha_3} \dots S_{r}^{\alpha_{r}}
V_{1,2}^{\eps} S_1^{2(\alpha_1+\alpha_2)} V_{1,2}^{\eps}
S_{r}^{\alpha_{r}} \dots S_3^{\alpha_2} \psi). \end{split}
\end{equation*}
Applying again Lemma \ref{lm:comm}, we find, for every $a>0$, a
constant $D (n,\eps,a)$ such that
\begin{equation*}
\begin{split}
V_{1,2}^{\eps} S_1^{2(\alpha_1+\alpha_2)} V^{\eps}_{1,2} \leq  D
(n,\eps,a) \, N^{a/n} \sum_{m=1}^{\alpha_1+\alpha_2}
S_1^{2(\alpha_1 +\alpha_2 - m + 2)} S_{2}^2 N^{m-1}\,.
\end{split}
\end{equation*}
Hence
\begin{equation}\label{eq:up7}
\begin{split}
\sum_{r=1}^{n-1} &\sum_{\balpha_{r} : |\balpha_{r}|=n} N^{r}
r(r-1) (\psi, V_{1,2}^{\eps} \, S_1^{2\alpha_1} \dots
S_{r}^{2\alpha_{r}} V_{1,2}^{\eps} \psi) \\ &\leq C_3 (n,\eps,a)
N^{a/n} \sum_{r=1}^{n-1} \sum_{\balpha_{r} : |\balpha_{r}|=n}
\sum_{m=1}^{\alpha_1+\alpha_2} N^{r+ m-1} (\psi,
S_1^{2(\alpha_1+\alpha_2 - m + 2)} S_2^{2} S_3^{2\alpha_3}
\dots S_{r}^{2\alpha_{r}} \psi) \\
&\leq C_3 (n,\eps,a) N^{a/n} \sum_{\ell=1}^{n+1}
\sum_{\balpha_{\ell} : |\balpha_{\ell}|=n+2} N^{\ell} (\psi,
S_1^{2\alpha_1} \dots S_{\ell}^{2\alpha_{\ell}} \psi) \, .
\end{split}
\end{equation}
Similarly to (\ref{eq:up6}), in order to control the terms in the
sum over $m$, we used the trivial bound
\[ N^{r+m-1} S_1^{2(\alpha_1 + \alpha_2-m +2)} S_2^2
S_3^{2\alpha_3} \dots S_{r}^{2\alpha_{r}} \leq N^{r+m-1}
S_1^{2(\alpha_1 + \alpha_2-m +2)} S_2^2 S_3^{2\alpha_3} \dots
S_{r}^{2\alpha_{r}} S_{r+1}^2 \dots S_{r+m-1}^2 \, ,\] and the
inequality $r+m-1 \leq r-1+\alpha_1 + \alpha_2 \leq r +
k-1-(\alpha_3 + \dots \alpha_{r}) \leq n+1$.

Inserting (\ref{eq:up4b}), (\ref{eq:up6}), and (\ref{eq:up7}) in
(\ref{eq:up4}) we find
\begin{equation*}
\begin{split}
\sum_{r=1}^{n-1} \sum_{\balpha_{r} : |\balpha_{r}|=n} N^{r}\,
\sum_{i<j} (\psi, V_{i,j}^{\eps}S_1^{2\alpha_1} &\dots S_{r}^{2
\alpha_{r}} V_{i,j}^{\eps} \psi) \\ &\leq \; C_4(n,\eps,a) \,
N^{a/n} \sum_{\ell=1}^{n+1} \sum_{\balpha_{\ell} :
|\balpha_{\ell}|=n+2} N^{\ell} (\psi, S_1^{2\alpha_1} \dots
S_{\ell}^{2\alpha_{\ell}} \psi)\,.
\end{split}
\end{equation*}
Combining the last bound and (\ref{eq:up3b}) with (\ref{eq:up3})
we find
\begin{multline*}
\sum_{r=1}^{n-1} \sum_{\balpha_{r} : |\balpha_{r}|=n} N^{r}\,
(\psi, \wt H_N \, S_1^{2\alpha_1} \dots S_{r}^{2 \alpha_{r}} \wt
H_N \psi) \\ \leq  C_5 (n,\eps,a) N^{a/n} \sum_{\ell=1}^{n+1}
\sum_{\balpha_{\ell} : |\balpha_{\ell}|=n+2} N^{\ell}\, (\psi,
S_1^{2\alpha_1} \dots S_{\ell}^{2\alpha_{\ell}} \psi)\, .
\end{multline*}
Defining $C(n+2,\eps,a) = C^n C_1 (n,\eps,a) +
C(n,\eps,a(n-1)/n)\cdot C_5(n,\eps,a)$, the last equation,
together with (\ref{eq:up2b}) and (\ref{eq:up1}), implies that
\begin{equation*}
\begin{split}
(\psi, \wt H_N^{n+2} \psi) \leq \; &C^{n+2} (\psi, (S_1^2 + \dots
+ S_N^2)^{n+2} \psi) \\ &+ C(n+2,a,\eps) N^{a} \sum_{\ell=1}^{n+1}
\sum_{\balpha_{\ell} : |\balpha_{\ell}|=n+2} N^{\ell}\, (\psi,
S_1^{2\alpha_1} \dots S_{\ell}^{2\alpha_{\ell}} \psi)
\end{split}
\end{equation*}
and completes the proof of the proposition.
\end{proof}

In order to apply the last proposition to our initial state
$\psi_N^{\kappa}$ we need the following lemma, which shows,
together with Proposition \ref{prop:upper}, that
\begin{equation}\label{eq:upbou}
 (\psi_N^{\kappa}, \wt H_N^k \psi_N^{\kappa} )
\leq C^k \end{equation} for $N$ large enough (depending on
$\eps,\kappa$ and $k$).

\begin{lemma}\label{lm:init}
Fix $\ph \in H^1 (\bR^3)$ with $\| \ph \| =1$. For $\kappa >0$
define $\ph^{\kappa} = \exp(-\kappa |p|/N) \ph$. Let
$\psi_N^{\kappa} (\bx) = \prod_{j=1}^N \ph^{\kappa} (x_j)$.
\begin{itemize}
\item[i)] We have $\| \ph^{\kappa} \| \leq 1$ and \[
(\ph^{\kappa}, (1 + p^2)^{k/2} \ph^{\kappa}) \leq C^k (1 + k! \,
(N/\kappa)^{k-2}) \| \ph \|_{H^1}^2 \, .\] \item[ii)] For every
$k\geq 1$ and $\kappa
>0$ there exists a constant $C(k,\kappa)$ such that
\begin{equation*}
\sum_{\ell=1}^{k-1} \sum_{\balpha_{\ell} : |\balpha_{\ell}|=k}
N^{\ell}\, (\psi_N^{\kappa}, S_1^{2\alpha_1} \dots
S_{\ell}^{2\alpha_{\ell}} \psi_N^{\kappa}) \leq C(k,\kappa)
N^{k-1} \, .
\end{equation*}
\item[iii)] For fixed $k$, and $\kappa$, there exists $N_0
(k,\kappa)$ such that
\begin{equation*}
(\psi_N^{\kappa}, (S_1^2 + \dots S_N^2)^k \psi_N^{\kappa}) \leq
C^k N^k
\end{equation*}
for all $N > N_0$ (the constant $C$ is independent of $k,\kappa$
and $N$).
\end{itemize}
\end{lemma}
\begin{proof}
i) The inequality $\| \ph^{\kappa} \| \leq 1$ follows from
$e^{-\kappa |p|/N} \leq 1$. Next we compute
\begin{equation*}
\begin{split}
(\ph^{\kappa} , (1 +p^2)^{k/2} \ph^{\kappa}) &\leq 2^{k/2} \int
\rd p \, (1+|p|^k) e^{-2\kappa |p|/N} |\widehat{\ph} (p)|^2 \\
&\leq 2^{k/2} \left( 1  + \sup_p \left( |p|^{k-2} e^{-2\kappa
|p|/N} \right) \right) \, \int \rd p \, (1 + |p|^2) |\widehat{\ph} (p)|^2 \\
&\leq C^k (1 + k! \, (N/\kappa)^{k-2} ) \, \| \ph \|_{H^1}^2 \, .
\end{split}
\end{equation*}

ii) For any $\ell \leq k-1$ and $\balpha_{\ell} = (\alpha_1,
\dots, \alpha_{\ell}) \in \bN^{\ell}_+$, with
$|\balpha_{\ell}|=k$, we have, using part i),
\begin{equation*}
\begin{split}
N^{\ell} (\psi_N^{\kappa},  S_{1}^{2\alpha_1} \dots
S_{\ell}^{2\alpha_{\ell}} \psi_N^{\kappa}) &\leq N^{\ell}
\prod_{j=1}^{\ell} (\ph^{\kappa}, ( 1 +  p^2)^{\alpha_j/2}
\ph^{\kappa}) \\ &\leq N^k  N^{\ell-k} \prod_{j=1}^{\ell}
C^{\alpha_j} (1 + \alpha_j! \, \kappa^{2- \alpha_j} N^{\alpha_j
-2}) \| \ph \|_{H^1}^2
\\ &\leq C^k N^k \| \ph \|_{H^1}^{2k}
\prod_{j=1}^{\ell} (N^{1-\alpha_j} + \alpha_j! \, \kappa^{2-
\alpha_j} N^{-1})\, .
\end{split}
\end{equation*}
Since $\alpha_j \geq 1$ for all $j=1,\dots ,\ell$, and since $\ell
\leq k-1$ there is at least one $j \in \{ 1,\dots ,\ell\}$ such
that $\alpha_j \geq 2$. Thus,
\begin{equation*}
\sum_{\ell=1}^{k-1} \sum_{\balpha_{\ell} : |\balpha_{\ell}|=k}
N^{\ell} (\psi_N^{\kappa},  S_{1}^{2\alpha_1} \dots
S_{\ell}^{2\alpha_{\ell}} \psi_N^{\kappa})  \leq C^k \| \ph
\|_{H^1}^{2k} N^{k-1} \prod_{j=1}^k (1+k! \kappa^{2-k})^k =
N^{k-1} C(k,\kappa) \, .
\end{equation*}

iii) By the permutation symmetry of $\psi_N^{\kappa}$, and by part
ii), we have
\begin{equation*}
\begin{split}
(\psi_N^{\kappa}, (S_1^2 + \dots + S_N^2)^k \psi_N^{\kappa}) &\leq
N^k (\psi_N^{\kappa}, S_1^2 \dots S_k^2 \psi_N^{\kappa}) +
\sum_{\ell=1}^{k-1} \sum_{\balpha_{\ell}:|\balpha_{\ell}|=k}
N^{\ell} (\psi_N^{\kappa}, S_{1}^{2\alpha_1} \dots
S_{\ell}^{2\alpha_{\ell}} \psi_N^{\kappa}) \\ &\leq N^k
(\ph^{\kappa}, (1+p^2)^{1/2} \ph^{\kappa})^k + C(k,\kappa)
N^{k-1}\\ &\leq C^k N^k \left(1 + \frac{C(k,\kappa)}{C^k}
N^{-1}\right)
\end{split}
\end{equation*}
where the constant $C$ only depends on $\| \ph \|_{H^1}$. For
fixed $k,\kappa$ we can now choose $N_0$ large enough such that
$C(k,\kappa) N^{-1} \leq 1$. Then for $N >N_0$ we have
$(\psi_N^{\kappa},(S_1^2 + \dots + S_N^2)^k \psi_N^{\kappa}) \leq
(2C)^k N^k$, which proves the claim.
\end{proof}

Using the bound (\ref{eq:upbou}), the conservation of the energy,
and the lower bound for $\wt H_N^k$ given in Proposition
\ref{prop:lower}, we finally arrive at the a-priori bound
(\ref{eq:ap}).

\begin{theorem}[A-Priori Estimate]\label{thm:apriori}
Fix a one-particle wave function $\ph \in H^1 (\bR^3)$, with $\|
\ph \| =1$. For $\kappa >0$ let $\ph^{\kappa} = \exp
(-\kappa|p|/N) \ph$. Define $\psi^{\kappa}_N (\bx) = \prod_{j=1}^N
\ph^{\kappa} (x_j)$. Let $\psi^{\kappa,\eps}_{N,t}$ be the
solution of the Schr\"odinger equation (\ref{eq:schrcut}) with
modified interaction and initial data $\psi_N^{\kappa}$. Then, for
all fixed $\eps, \kappa
>0$ and $k\geq 1$ there exists $N_0 (\eps,\kappa,k)$ such that
\begin{equation}\label{eq:deriv}
(\psi_{N,t}^{\eps,\kappa} , \, S_1^2 \dots S_k^2 \, \,
\psi_{N,t}^{\eps, \kappa}) \leq C^k \quad \quad \text{and} \quad
\quad (\psi_{N,t}^{\eps,\kappa}, \, S_1^4 S_2^2 \dots S_{k-1}^2 \,
\, \psi_{N,t}^{\eps,\kappa}) \leq C^k \, N
\end{equation}
for all $N \geq N_0$ and for all $t\in \bR$. The constant $C$ only
depends on $\lambda$ and $\| \ph \|_{H^1}$. Denoting by $\wt
\gamma_{N,t}^{(k)}$ the $k$-particle marginal distribution
associated to $\psi^{\eps,\kappa}_{N,t}$, (\ref{eq:deriv}) is
equivalent to
\begin{equation*}
\tr \; \left| S_1 \dots S_k \, \wt \gamma^{(k)}_{N,t} \, S_k \dots
S_1 \, \right| \leq C^k \quad \quad \text{and} \quad \quad \tr \;
\left| S_1^2 S_2 \dots S_{k-1} \, \wt \gamma^{(k)}_{N,t} S_{k-1}
\dots S_2 S_1^2 \right| \leq C^k N
\end{equation*}
for all $N \geq N_0$.
\end{theorem}

\begin{proof}
{F}rom Proposition \ref{prop:upper} we have
\begin{equation*}
(\psi_N^{\kappa}, \wt H_N^k \psi_{N}^{\kappa}) \leq C^k
(\psi_{N}^{\kappa}, (S_1^2 + \dots + S_N^2)^k \psi_{N}^{\kappa}) +
C(k,\eps,a) N^a \sum_{\ell=1}^{k-1} \sum_{\balpha_{\ell} :
|\balpha_{\ell}|=k} N^{\ell}\, (\psi_N^{\kappa}, S_1^{2\alpha_1}
\dots S_{\ell}^{2\alpha_{\ell}} \psi_N^{\kappa})\,.
\end{equation*}
By Lemma \ref{lm:init}, part ii) and iii) we find
\begin{equation*}
(\psi_N^{\kappa}, \wt H^k_N \psi_{N}^{\kappa}) \leq C^k N^k +
C(k,\eps,\kappa,a) N^a N^{k-1} \leq C^k N^k \left( 1 +
\frac{C(k,\eps,\kappa,a)}{C^k} N^{a-1}\right)
\end{equation*}
for $N$ large enough, depending on $k$ and $\kappa$. Hence, fixing
$a<1$, we find
\begin{equation*}
(\psi_N^{\kappa}, \wt H^k_N \psi_{N}^{\kappa}) \leq C_1^k N^k
\end{equation*}
for all $N$ large enough, depending on $k,\eps,\kappa$, and with a
constant $C_1$ only depending on $\| \ph \|_{H^1}$ and $\lambda$.
By the conservation of the energy and by Proposition
\ref{prop:lower}, we get
\begin{equation*}
\begin{split}
C_1^k N^k &\geq (\psi_N^{\kappa}, \, \wt H^k_N \,
\psi_{N}^{\kappa}) = (\psi_{N,t}^{\eps,\kappa}, \, \wt H^k_N \,
\psi_{N,t}^{\eps,\kappa})
\\ &\geq C_2^k N^k (\psi_{N,t}^{\eps,\kappa}, \, S_1^2 \dots S_k^2
\, \psi_{N,t}^{\eps,\kappa}) + C_2^k N^{k-1}
(\psi_{N,t}^{\eps,\kappa}, \, S_1^4 S_2^2 \dots S_{k-1}^2 \,
\psi_{N,t}^{\eps,\kappa}) \end{split}
\end{equation*} for all $N$
large enough.
\end{proof}

\section{Compactness of the sequence $\wt \Gamma_{N,t}$.}
\label{sec:compact}\setcounter{equation}{0}

Recall that we defined $\wt \Gamma_{N,t} = \{ \wt
\gamma^{(k)}_{N,t} \}_{k=1}^N$ as the marginal densities
associated to the solution $\psi_{N,t}^{\eps,\kappa}$ of the
regularized Sch\"odinger equation (\ref{eq:schrcut}) with initial
data $\psi_N^{\kappa}$. They satisfy the regularized BBGKY
hierarchy
\begin{equation}\label{eq:BBGKYcut}
\begin{split}
i\partial_t \wt\gamma^{(k)}_{N,t}  = \; &\sum_{j=1}^k \,
\left[(1-\Delta_{x_j})^{1/2}, \wt \gamma_{N,t}^{(k)} \right] +
\lambda N^{-1} \sum_{1 \leq i<j \leq k} \left[ V^{\eps} (x_i -
x_j), \wt\gamma^{(k)}_{N,t}\right] \\ &+\lambda (1 -\frac{k}{N})
\sum_{j=1}^k \tr_{k+1} \left[ V^{\eps} (x_j -x_{k+1}),
\wt\gamma_{N,t}^{(k+1)}\right] \, ,
\end{split}
\end{equation}
where $V^{\eps} (x) = (|x| + \eps N^{-1})^{-1}$.

\begin{theorem}\label{thm:compact}
Fix $k \geq 1$ and $T \geq 0$. Then $\wt \gamma_{N,t}^{(k)} \in
C([0,T], \cH_k)$ for all $N$ large enough (depending on $k$,
$\eps$ and $\kappa$). Moreover the sequence $\wt
\gamma_{N,t}^{(k)}$ is compact in $C([0,T],\cH_k)$ with respect to
the metric $\widehat \rho_k$ (defined in (\ref{eq:hatrho})), and,
if $\gamma^{(k)}_{\infty,t} \in C([0,T], \cH_k)$ denotes an
arbitrary limit point of the sequence $\wt \gamma^{(k)}_{N,t}$
with respect to the metric $\widehat \rho_k$, we have
\begin{equation}\label{eq:aprioriboun}
\| \gamma_{\infty,t}^{(k)} \|_{\cH_k} = \tr \; |S_1 \dots S_k
\gamma^{(k)}_{\infty,t} S_k \dots S_1| \leq C^k
\end{equation}
for all $t \in [0,T]$ (the constant $C$ is independent of
$k,\eps$, and $\kappa$).
\end{theorem}
\begin{proof}
We prove that the sequence $\wt \gamma^{(k)}_{N,t}$ is
equicontinuous in $t$ with respect to the metric $\wh \rho_k$. To
this end we rewrite the BBGKY hierarchy (\ref{eq:BBGKYcut}) in the
integral form
\begin{equation*}
\begin{split}
\wt \gamma^{(k)}_{N,t} = \; &\wt \gamma^{(k)}_{N,s} - i
\sum_{j=1}^k \int_s^t \rd \tau \, [ S^2_j , \wt
\gamma^{(k)}_{N,\tau}] -i\frac{\lambda}{N} \sum_{i<j}^k \int_s^t
\rd \tau \, [V^{\eps} (x_i -x_j), \wt \gamma^{(k)}_{N,\tau}]
\\ &-i (1 - \frac{1}{N}) \lambda \sum_{j=1}^k \int_s^t \rd \tau \,
\tr_{k+1} [ V^{\eps} (x_j -x_{k+1}), \wt \gamma^{(k+1)}_{N,\tau}]
\end{split}
\end{equation*}
where $\tr_{k+1}$ denotes the partial trace over the $k+1$'th
variable. Next we choose $J^{(k)} \in \cK_k$ such that $\| S_j
J^{(k)} S_j^{-1} \|$ is bounded for all $j=1,\dots k$, and we
compute
\begin{equation*}
\begin{split}
\tr^{(k)} &\; J^{(k)} \left( \wt\gamma^{(k)}_{N,t} -
\wt\gamma^{(k)}_{N,s}\right)  = - i \sum_{j=1}^k \int_s^t \rd \tau
\, \tr^{(k)} \; \left( S_j^{-1} J^{(k)} S_j - S_j J^{(k)} S_j^{-1}
\right) \, S_j \wt \gamma^{(k)}_{N,\tau} S_j \\ &- i
\frac{\lambda}{N} \sum_{i<j}^k \int_s^t \rd \tau \, \tr^{(k)}
\;\left( S_j^{-1} J^{(k)} S_j S_j^{-1} V^{\eps} (x_i -x_j)
S_j^{-1} S_j \wt \gamma^{(k)}_{N,\tau} S_j \right. \\
&\hspace{5cm}\left. - S_j J^{(k)} S_j^{-1} S_j \wt
\gamma^{(k)}_{N,\tau} S_j S_j^{-1} V^{\eps} (x_i - x_j) S_j^{-1}
\right) \\ &-i (1 - \frac{1}{N}) \lambda \sum_{j=1}^k \int_s^t \rd
\tau \, \tr^{(k+1)} J^{(k)} [ S_{k+1}^{-1} V(x_j -x_{k+1})
S_{k+1}^{-1}, S_{k+1} \wt \gamma^{(k)}_{N,\tau}S_{k+1}]
\end{split}
\end{equation*}
where we used the fact that $S_{k+1} = (1 + p_{k+1}^2)^{1/4}$
commutes with $J^{(k)}$. Since we assumed $\| S_j J^{(k)} S_j^{-1}
\|$ to be bounded for all $j=1,\dots k$ and since $\| S_j^{-1}
V^{\eps} (x_i -x_j) S_j^{-1} \| \leq C$ (see Lemma
\ref{lm:norm-bound}), we find, using Theorem \ref{thm:apriori},
\begin{equation*}
\left|\tr^{(k)} \; J^{(k)} \left(\wt  \gamma^{(k)}_{N,t} - \wt
\gamma^{(k)}_{N,s}\right)\right|  \leq C_k \, |t-s|
\end{equation*}
for a constant $C_k$ which depends on $k$ and $J^{(k)}$, but is
independent of $t,s \in [0,T]$ and of $N$, for $N$ large enough
(depending on $k$, $\eps$ and $\kappa$). Since the set of all
$J^{(k)} \in \cK_k$, with $\| S_j J^{(k)} S_j^{-1} \|$ bounded for
$j=1,\dots k$, is a dense subset of $\cA_k$, it follows by Lemma
\ref{lm:equicont} that the sequence $\wt \gamma_{N,t}^{(k)}$ is
equicontinuous with respect to the metric $\rho_k$ (defined in
(\ref{eq:rho})). This in particular implies that $\wt
\gamma_{N,t}^{(k)} \in C([0,T], \cH_k)$ for all $N$ large enough.
Moreover, since the sequence $\gamma_{N,t}^{(k)} \in C([0,T],
\cH_k)$ is uniformly bounded, it follows by the Arzela-Ascoli
Theorem (see \cite{RS}, Theorem I.28) that it is compact with
respect to the metric $\wh \rho_k$, and that any limit point
$\gamma^{(k)}_{\infty,t}$ satisfies (\ref{eq:aprioriboun}).
\end{proof}

\section{Convergence to the infinite hierarchy}
\label{sec:conv}\setcounter{equation}{0}

In this section we consider limit points of the sequence $\wt
\Gamma_{N,t} = \{ \wt \gamma_{N,t}^{(k)} \}_{k=1}^N$, where $\wt
\gamma_{N,t}^{(k)}$ are the marginals associated with the solution
$\psi_{N,t}^{\eps,\kappa}$ of the modified Schr\"odinger equation
(\ref{eq:schrcut}) with regularized initial data
$\psi_N^{\kappa}$. We prove that any limit point of this sequence,
as $N \to \infty$, is a solution of the infinite hierarchy of
equations (\ref{eq:BBGKYinf}).

\begin{theorem}\label{thm:convergence}
Suppose $\Gamma_{\infty,t} = \{ \gamma^{(k)}_{\infty,t} \}_{k\geq
1} \in C([0,T], \cH)$ is a limit point of the sequence $\wt
\Gamma_{N,t} = \{ \wt \gamma^{(k)}_{N,t} \}_{k = 1}^N$ with
respect to the product of the topologies induced by the metrics
$\wh \rho_k$ (defined in (\ref{eq:hatrho})). Then
$\Gamma_{\infty,t}$ satisfies the infinite hierarchy
(\ref{eq:BBGKYinf}) with initial data
\begin{equation}\label{eq:initial}
\gamma^{(k)}_{\infty,0} (\bx_k ; \bx'_k) = \gamma_0^  {(k)} (\bx_k
;\bx'_k)= \prod_{j=1}^k \ph (x_j) \overline{\ph (x_j)} \, .
\end{equation}
\end{theorem}

\begin{proof}
Without loss of generality we can assume that $\wt \Gamma_{N,t}
\to \Gamma_{\infty,t} \in C([0,T], \cH)$ with respect to the
product of the topologies induced by the metrics $\wh \rho_k$.
Then, for every fixed $k \geq 1$ and $t \in [0,T]$, we have $\wt
\gamma_{N,t}^{(k)} \to \gamma_{\infty,t}^{(k)} \in \cH_k$ with
respect to the weak* topology of $\cH_k$ (because the sequence
$\gamma_{N,t}^{(k)}$ is uniformly bounded in $\cH_k$ and the
metric $\rho_k$ is equivalent to the weak* topology for uniformly
bounded sequences). Hence, for every $t \in [0,T]$, $k \geq 1$,
and every $J^{(k)} \in \cA_k$, we have
\begin{equation}\label{eq:converg}
\tr \; J^{(k)} \left( \wt \gamma^{(k)}_{N,t} -
\gamma_{\infty,t}^{(k)} \right) \to 0\quad \text{for } N \to
\infty \, .\end{equation} We choose $J^{(k)} \in \cK_k$ such that
$S_j J^{(k)} S_j^{-1}$ is bounded for all $j=1,\dots k$, and we
obtain, from the regularized BBGKY hierarchy (\ref{eq:BBGKYcut}),
rewritten in integral form,
\begin{equation}\label{eq:BBGKYinte}
\begin{split}
\tr^{(k)} \; J^{(k)} \wt\gamma^{(k)}_{N,t} = \; &\tr^{(k)} J^{(k)}
\cU^{(k)} (t) \wt \gamma^{(k)}_{N,0} \\ &- i \lambda (1-k/N)
\sum_{j=1}^k \int_0^t \rd s \, \tr^{(k+1)} \; J^{(k)} \cU^{(k)}
(t-s) \; [V^{\eps}_{j,k+1} , \wt \gamma^{(k+1)}_{N,s}]
\end{split}
\end{equation} where
\[ \cU^{(k)} (t) \gamma^{(k)} = e^{-i t  H^{(k)}}
\gamma^{(k)} e^{it H^{(k)}} \quad \quad \text{with } \quad H^{(k)}
= \sum_{j=1}^k S_j^2 + \frac{\lambda}{N} \sum_{i<j}^k
V^{\eps}_{i,j} \] and where we use the notation $V^{\eps}_{i,j} =
V^{\eps} (x_i - x_j)$. Next we consider the limit $N \to \infty$
of (\ref{eq:BBGKYinte}). {F}rom (\ref{eq:converg}), the l.h.s. of
(\ref{eq:BBGKYinte}) converges to $\tr \; J^{(k)}
\gamma_{\infty,t}^{(k)}$. As for the first term on the r.h.s. of
(\ref{eq:BBGKYinte}) we find
\begin{equation}\label{eq:conv1}
\begin{split}
\tr^{(k)} \, J^{(k)} \, \left( \cU^{(k)} (t) \wt
\gamma^{(k)}_{N,0} - \cU_0^{(k)} (t) \gamma^{(k)}_{\infty,0}
\right) = \; &\tr^{(k)} \, J^{(k)} \, \left( \cU^{(k)} (t) -
\cU_0^{(k)} (t) \right)\wt \gamma^{(k)}_{N,0} \\ &+ \tr^{(k)} \;
J^{(k)} \, \cU_0^{(k)} (t) \left( \wt \gamma^{(k)}_{N,0}-
\gamma^{(k)}_{\infty,0} \right)
\end{split}
\end{equation}
where we used the notation
\[ \cU^{(k)}_0 (t) \gamma^{(k)} = \exp \left( -it\sum_{j=1}^k S_j^2 \right)
\gamma^{(k)} \exp \left(it\sum_{j=1}^k S_j^2 \right) \] for the
free evolution of the first $k$ particles. The first contribution
on the r.h.s. of (\ref{eq:conv1}) can be handled as follows:
\begin{equation*}
\tr^{(k)} \, J^{(k)} \, \left( \cU^{(k)} (t) - \cU_0^{(k)} (t)
\right) \wt \gamma^{(k)}_{N,0}=\frac{-i\lambda}{N} \sum_{i,j}^k
\int_0^t \rd s \, \tr^{(k)} \, J^{(k)} \cU^{(k)} (t-s)
V^{\eps}_{i,j} \, \cU_0^{(k)} (s) \wt \gamma_{N,0}^{(k)}
\end{equation*}
and thus, using the permutation symmetry of $\wt
\gamma_{N,0}^{(k)}$, and the a-priori estimate from Theorem
\ref{thm:apriori},
\begin{equation*}
\begin{split}
\left|\tr^{(k)} \, J^{(k)} \, \left( \cU^{(k)} (t) - \cU_0^{(k)}
(t) \right) \wt \gamma^{(k)}_{N,0}\right| &\leq C N^{-1} \, t \,
k^2 \| J^{(k)} \|  \, \| V^{\eps}_{1,2} S_1^{-1} S_2^{-1}\| \,
\tr^{(k)} \, S_1 S_2 \wt \gamma_{N,0}^{(k)} S_2 S_1 \\& \leq C
N^{-1}
\end{split}
\end{equation*}
with a constant $C$ depending on $k$, on the observable $J^{(k)}$
and on $t$. As for the second term on the r.h.s. of
(\ref{eq:conv1}) we have
\begin{equation*}
\tr^{(k)} \, J^{(k)} \, \cU_0^{(k)} (t) \left(\wt
\gamma^{(k)}_{N,0}- \gamma^{(k)}_{\infty,0} \right) = \tr^{(k)} \,
\left(\cU_0^{(k)} (-t) J^{(k)} \right) \, \left(\wt
\gamma^{(k)}_{N,0} - \gamma^{(k)}_{\infty,0} \right) \to 0
\end{equation*}
for $N \to \infty$, because, if $J^{(k)} \in \cK_k$, then also
$\cU_0^{(k)} (-t) J^{(k)} \in \cK_k \subset \cA_k$, and thus
(\ref{eq:converg}) can be used.

Next we prove that the difference
\begin{multline*}
(1-k/N) \sum_{j=1}^k \int_0^t \rd s  \, \tr^{(k+1)} \; J^{(k)}
\cU^{(k)} (t-s) \, [V^{\eps}_{j,k+1} , \wt \gamma^{(k+1)}_{N,s} ]
\\ -  \sum_{j=1}^k \int_0^t \rd s \, \tr^{(k+1)} \; J^{(k)}\,
\cU_0^{(k)} (t)\, [V_{j,k+1} , \gamma^{(k+1)}_{\infty,s} ]
\end{multline*}
converges to zero, for $N \to \infty$. Here and henceforth,
$V_{i,j} = V (x_i -x_j) = |x_i -x_j|^{-1}$. To this end, we
rewrite it as the sum of four terms
\begin{equation*}\label{eq:difference}
\begin{split}
&- k/N \sum_{j=1}^k \int_0^t \rd s \tr^{(k+1)} \; J^{(k)} \,
\cU^{(k)} (t) \, [ V^{\eps}_{j,k+1}, \wt \gamma^{(k)}_{N,s}]
\\ &+ \sum_{j=1}^k \int_0^t \rd s \, \tr^{(k+1)}
\; J^{(k)} \, \left( \cU^{(k)} (t-s) - \cU_0^{(k)} (t-s) \right) [
V^{\eps}_{j,k+1}, \wt \gamma^{(k)}_{N,s}]
\\ & +\sum_{j=1}^k \int_0^t \rd s \, \tr^{(k+1)} \; J^{(k)} \,
\cU_0^{(k)} (t-s) [ V^{\eps}_{j,k+1} - V_{j,k+1}, \wt
\gamma^{(k)}_{N,s}]
\\ &+ \sum_{j=1}^k \int_0^t \rd s
\,\tr_{k+1}\; J^{(k)} \,\cU_0^{(k)} (t-s)  \left[ V_{j,k+1},
\left(\wt \gamma^{(k)}_{N,s} - \gamma_{\infty,s}^{(k)}\right)
\right]\, .
\end{split}
\end{equation*}
The first term converges to zero, for $N \to \infty$, because
\begin{equation*}
\begin{split}
&\left|k/N \sum_{j=1}^k \int_0^t \rd s \, \tr^{(k+1)} \; J^{(k)}
\, \cU^{(k)} (t) \, [ V^{\eps}_{j,k+1}, \wt
\gamma^{(k+1)}_{N,s}]\right| \\ &\hspace{3cm} \leq C \,N^{-1} k^2
t \, \| J^{(k)} \| \, \| V^{\eps}_{1,k+1} S_1^{-1} S_{k+1}^{-1} \|
\, \sup_{s\in [0,t]} \tr^{(k+1)} \; |S_1 S_{k+1}
\gamma^{(k+1)}_{N,s}|
\\ &\hspace{3cm} \leq C_{k,t} N^{-1} \, \sup_{s \in [0,t]}
\tr^{(k+1)} \; S_1 S_{k+1} \gamma^{(k+1)}_{N,s} S_{k+1} S_1 \leq
C_{k,t} N^{-1}
\end{split}
\end{equation*}
for a constant $C_{k,t}$ depending on $k$, on $t$, and on the
observable $J^{(k)}$. To control the second term we note that
\begin{equation*}
\begin{split}
\sum_{j=1}^k & \int_0^t \rd s \, \tr^{(k+1)} \; J^{(k)} \, \left(
\cU^{(k)} (t-s) - \cU_0^{(k)} (t-s) \right) \left[
V^{\eps}_{j,k+1}, \wt \gamma^{(k+1)}_{N,s} \right] \\ &= -i\lambda
N^{-1} \sum_{j=1}^k  \sum_{\ell<m}^k \int_0^t \rd s \int_0^{t-s}
\rd \tau \, \tr^{(k+1)} \, J^{(k)} \, \cU_0^{(k)} (t-s-\tau)
\left[ V^{\eps}_{\ell,m}, \cU^{(k)} (r) \left[ V^{\eps}_{j,k+1},
\wt \gamma^{(k+1)}_{N,s}\right]\right] \,.
\end{split}
\end{equation*}
Writing down the four terms arising from the two commutators and
using the permutation symmetry, we get the bound
\begin{equation*}
\begin{split}
&\left| \sum_{j=1}^k  \int_0^t \rd s \, \tr^{(k+1)} \; J^{(k)} \,
\left( \cU^{(k)} (t-s) - \cU_0^{(k)}
(t-s) \right) \left[ V^{\eps}_{j,k+1}, \wt \gamma^{(k+1)}_{N,s} \right]\right| \\
&\hspace{2cm}\leq C t^2 k^3 N^{-1} \, \sup_{\ell \leq k} \left(\|
S_{\ell} J^{(k)} S_{\ell}^{-1}\| + \| J^{(k)} \| \right) \, \|
S_{1}^{-1} V^{\eps}_{1,2} \| \| V^{\eps}_{1,k+1} S_1^{-1}
S_{k+1}^{-1} \| \\ &\hspace{6cm} \times \sup_{s \in [0,T]} \left(
\tr \; |S_1 S_2 \wt \gamma^{(k+1)}_{N,s} S_3| + \tr\, |S_1 S_2 S_3
\wt \gamma^{(k+1)}_{N,s}| \right)
 \\
&\hspace{2cm} \leq C_{\eps,t,k} N^{-1/2} \, \sup_{s \in [0,T]} \,
\tr \, S_1 S_2 S_3 \wt \gamma^{(k+1)}_{N,s} S_1 S_2 S_3 \leq C
N^{-1/2}
\end{split}
\end{equation*}
where the constant $C$ depends on $\eps, k, t$ and on $J^{(k)}$.
Here we used that $\| S_{\ell}^{-1} V^{\eps}_{\ell,m} \| \leq C
\eps^{-1/2} N^{1/2}$, the symmetry of $\wt \gamma^{(k+1)}_{N,s}$,
and Theorem \ref{thm:apriori}. To bound the third term in
(\ref{eq:difference}), we note that
\begin{equation*} V_{j,k+1}^{\eps} - V_{j,k+1} = \frac{\eps
N^{-1}}{|x_j -x_{k+1}| (|x_j -x_{k+1}| + \eps N^{-1})}.
\end{equation*} Hence
\begin{equation*}
\begin{split}
&\left| \sum_{j=1}^k \int_0^t \rd s \, \tr^{(k+1)} \; J^{(k)} \,
\cU_0^{(k)} (t-s) \left[ V^{\eps}_{j,k+1} - V_{j,k+1}, \wt
\gamma^{(k+1)}_{N,s}\right]\right| \\ &\hspace{1cm} \leq k t \eps
N^{-1} \, \| S_1^{-1} J^{(k)} S_1 \| \,  \| S_1^{-1}
\frac{1}{|x_1- x_2|(|x_1 - x_2| + \eps N^{-1})} S_1^{-1} S_2^{-2}
\| \, \sup_{s \in [0,T]} \tr \; |S_1 S_2 \wt \gamma^{(k+1)}_{N,s}
S_1| \\ &\hspace{1cm} \leq C_{k,t} N^{-1/2} \, \sup_{s \in [0,T]}
\tr \; S_1 S_2 \wt \gamma^{(k+1)}_{N,s} S_1 S_2 \leq C N^{-1/2}
\end{split}
\end{equation*}
for a constant $C_{k,t}$ depending on $k$, $t$, and $J^{(k)}$.
Finally we consider the last term in (\ref{eq:difference})
\begin{equation}\label{eq:fourth}
\begin{split}
\sum_{j=1}^k  \int_0^t \rd s \,\tr_{k+1} \; J^{(k)} \,\cU_0^{(k)}
(t-s)  \left[ V_{j,k+1}, \left(\wt \gamma^{(k+1)}_{N,s} -
\gamma_{\infty,s}^{(k+1)}\right) \right]
\end{split}
\end{equation}
This term converges to zero, because, since $J^{(k)} \in \cK_k
\subset \cA_k$, we have $(\cU_0^{(k)} (s-t) J^{(k)}) V_{j,k+1} \in
\cA_{k+1}$ for every $j \leq k$, every $s$ and $t$. In fact
\begin{multline*}
\| S_1^{-1} \dots S_{k+1}^{-1} (\cU_0^{(k)} (s-t) J^{(k)})
V_{j,k+1} S_1^{-1} \dots S_{k+1}^{-1} \| \\ \leq \| S_1^{-1} \dots
S_k^{-1} J^{(k)} S_k^{-1} \dots S_1^{-1} \| \, \| S_{k+1}^{-1}
V_{j,k+1} S_{k+1}^{-1} \| \leq C \| J^{(k)} \|_{\cA_k}.
\end{multline*}
This proves that the integrand in (\ref{eq:fourth}) converges to
zero as $N \to \infty$, for every $s \in [0,t]$ and for every
$j=1,\dots ,k$. Since the integrand is uniformly bounded in $s$,
it follows that (\ref{eq:fourth}) converges to zero for $N \to
\infty$.

We have proven that, for every fixed $t \in [0,T]$, $k \geq 1$ and
$J^{(k)} \in \cK_k$ such that $S_j J^{(k)} S_j^{-1}$ is finite for
all $j$, we have
\begin{equation}\label{eq:conv5}
\tr^{(k)} \, J^{(k)} \gamma^{(k)}_{\infty,t} = \tr \, J^{(k)}
\cU_0^{(k)} (t) \gamma^{(k)}_{\infty,0} -i\lambda \sum_{j=1}^k
\int_0^t \rd s \, \tr^{(k+1)}\, J^{(k)} \cU_0^{(k)} (t-s) \left[
V_{j,k+1}, \gamma^{(k+1)}_{\infty,s} \right]\,.
\end{equation}
Since the set of $J^{(k)} \in \cK_k$ such that $\sup_{j\leq k} \|
S_j^{-1} J^{(k)} S_j\| < \infty$ is a dense subset of $\cA_k$, it
follows by a simple approximation argument, that (\ref{eq:conv5})
holds true for all $J^{(k)} \in \cA_k$. Thus
\begin{equation*}
\gamma^{(k)}_{\infty,t} = \cU_0^{(k)} (t) \gamma_{\infty,0}^{(k)}
-i\lambda \sum_{j=1}^k \int_0^t \rd s \, \cU_0^{(k)} (t-s) \,
\tr_{k+1} \left[ V_{j,k+1}, \gamma_{\infty,s}^{(k+1)} \right]
\end{equation*}
for all $t \in [0,T]$. Finally we have to prove that
$\gamma_{\infty,0}^{(k)}$ is given by (\ref{eq:initial}). Recall
that \[ \wt \gamma_{N,0}^{(k)} (\bx_k;\bx'_k) =\prod_{j=1}^k
\ph^{\kappa} (x_j) \overline{\ph^{\kappa}} (x'_j) \] with
$\ph^{\kappa} = \exp (-\kappa |p|/N) \ph$. Hence, for any $J^{(k)}
\in \cK_k$, we have
\begin{equation}\label{eq:conv6}
\tr \, J^{(k)} \left( \gamma_{0}^{(k)} - \gamma_{\infty,0}^{(k)}
\right)=\tr \, J^{(k)} \left( \gamma_{0}^{(k)} - \wt
\gamma_{N,0}^{(k)} \right) + \tr \, J^{(k)} \left( \wt
\gamma_{N,0}^{(k)} - \gamma_{\infty,0}^{(k)} \right)\,.
\end{equation} The second term converges to zero, for $N \to
\infty$. As for the first one, we have
\begin{equation*}
\left| \tr \, J^{(k)} \left( \gamma_0^{(k)} - \wt
\gamma_{N,0}^{(k)} \right) \right| \leq \| J^{(k)} \| \, \tr
\left| \gamma_0^{(k)} - \wt \gamma_{N,0}^{(k)}\right| \leq C k \,
\|\ph - \ph^{\kappa} \| \leq C k N^{-1}
\end{equation*}
where we used that $\| \ph^{\kappa} \|\leq \| \ph \| =1$, and that
\[ \| \ph - \ph^{\kappa} \|^2 = \int \rd p \, (1-
e^{-\kappa|p|/N})^2 |\ph (p)|^2 \leq \kappa^2 N^{-2} \| \ph
\|_{H^1}\,. \] Since the choice of $N$ on the right side of
(\ref{eq:conv6}) is arbitrary, we find
\begin{equation*}
\tr \, J^{(k)}\left( \gamma_{0}^{(k)} - \gamma_{\infty,0}^{(k)}
\right)= 0
\end{equation*}
for all $J^{(k)} \in \cK_k$. Moreover, using the fact that
$\gamma_0^{(k)}$ and $\gamma_{\infty,0}^{(k)}$ have finite $\cH_k$
norm, a simple approximation argument shows that the last equation
is true for all $J^{(k)} \in \cA_k$. This proves that
$\gamma_{\infty,0}^{(k)} = \gamma_0^{(k)}$.
\end{proof}

\section{Uniqueness of the solution of the infinite hierarchy}
\label{sec:unique}\setcounter{equation}{0}

In this section we prove the uniqueness of the solution of the
infinite hierarchy (\ref{eq:BBGKYinf}). We already know that
$\Gamma_t = \{ \gamma_t^{(k)} \}_{k \geq 1}$, with
\[ \gamma_t^{(k)} (\bx_k;\bx'_k) = \prod_{j=1}^k \ph_t (x_j)
\overline{\ph_t} (x_j)\] and $\ph_t$ the solution of the nonlinear
Hartree equation (\ref{eq:hartree}), is a solution of
(\ref{eq:BBGKYinf}). Since, by Theorem \ref{thm:convergence}, we
know that every limit point of $\wt \Gamma_{N,t}$ is a solution of
(\ref{eq:BBGKYinf}), it follows that $\Gamma_t$ is the only limit
point of $\wt \Gamma_{N,t}$.

\begin{theorem}\label{thm:unique}
Fix $T>0$. Let $\Gamma_{0} = \{ \gamma^{(k)}_{0}\}_{k \geq 1} \in
\cH=\bigoplus_{k \geq 1} \cH_k$. Moreover, suppose that there
exists $C
>0$ such that
\[ \| \gamma^{(k)}_0 \|_{\cH_k} = \tr \; |S_1 \dots S_k \gamma^{(k)}_0 S_k \dots S_1|
\leq C^k \, \] for all $k \geq 1$. Then there exists at most one
solution $\Gamma_{t} \in C([0,T], \cH)$ of the infinite hierarchy
(\ref{eq:BBGKYinf}) such that $\Gamma_{t=0} = \Gamma_0$ and
\[ \| \gamma^{(k)}_t \|_{\cH_k} \leq C^k \]
for all $t \in [0,T]$.
\end{theorem}
\begin{proof}
Suppose $\Gamma_t = \{ \gamma^{(k)}_t \}_{k \geq 1}$ is a solution
of the infinite BBGKY hierarchy (\ref{eq:BBGKYinf}), so that
$\Gamma_{t=0} = \Gamma_0$ and \begin{equation}\label{eq:apri} \|
\gamma^{(k)}_t \|_{\cH_k} \leq C^k \end{equation} for all $t \in
[0,T]$ and $k \geq 1$. Rewriting (\ref{eq:BBGKYinf}) in integral
form we find
\begin{equation}\label{eq:hier}
\gamma^{(k)}_t = \cU^{(k)}_0 (t) \gamma_0^{(k)} -i \lambda
\sum_{j=1}^k \int_0^{t} \rd s \, \cU_0^{(k)} (t-s) \, \tr_{k+1} \,
[V_{j,k+1}, \gamma^{(k+1)}_{s}] \, ,
\end{equation}
where we use the notation $V_{i,j} = V(x_i -x_j) = |x_i
-x_j|^{-1}$ and where the free evolution $\cU_0^{(k)} (t)$ is
defined by
\begin{equation*}
\cU_0^{(k)} (t) \gamma^{(k)} = \exp \left(-it \sum_{j=1}^k
S_j^2\right) \gamma^{(k)} \exp \left(it \sum_{j=1}^k S^2_j\right)
\, .
\end{equation*}
We iterate (\ref{eq:hier}), and find
\begin{equation}\label{eq:zet}
\gamma^{(k)}_t = \cU_0^{(k)} (t) \gamma_0^{(k)} + \sum_{m=1}^{n-1}
\zeta (k,t,m) + \wt \zeta (k,t,n)\,
\end{equation}
where
\begin{equation*}
\begin{split}
\zeta (k,t,m) = &\; (-i\lambda)^m \sum_{j_1=1}^k \dots
\sum_{j_m=1}^{k+m-1} \int_0^t \rd s_1 \dots \int_0^{s_{m-1}} \rd
s_m \,  \cU_0^{(k)} (t-s_1) \, \tr_{k+1} \, \left[ V_{j_1,k+1},
\cU_0^{(k+1)} (s_1 -s_2) \right. \\ &\dots \times \cU^{(k+m-1)}_0
(s_{m-1}-s_m) \, \tr_{k+m} \, \left. \left[ V_{j_m,k+m},
\cU_0^{(k+m)} (s_m) \gamma_0^{(k+m)} \right] \dots \right]
\end{split}
\end{equation*}
and the error term $\wt \zeta (k,n,t)$ is given by
\begin{equation}\label{eq:wtzeta}
\begin{split}
\wt \zeta (k,t,n) = &\; (-i\lambda)^n \sum_{j_1=1}^k \dots
\sum_{j_n=1}^{k+n-1} \int_0^t \rd s_1 \dots \int_0^{s_{n-1}} \rd
s_n \,  \cU_0^{(k)} (t-s_1) \, \tr_{k+1} \, \left[V_{j_1,k+1},
\cU_0^{(k+1)} (s_1 -s_2) \right. \\ &\left. \dots
\times\cU_0^{(k+n-1)} (s_{n-1}-s_n) \, \tr_{k+n} \, \left[
V_{j_n,k+n}, \gamma_{s_n}^{(k+n)} \right] \dots \right] \, .
\end{split}
\end{equation}
In order to bound the error term, we note that, for any $\ell \geq
1$, $j=1,\dots \ell$, $s\in \bR$, and $\gamma^{(\ell+1)} \in
\cH_{\ell+1}$, we have
\begin{equation}\label{eq:gammaell}
\begin{split}
\| \cU_0^{(\ell)} (s) \tr_{\ell+1} [ V_{j,\ell+1} ,
&\gamma^{(\ell+1)} ] \|_{\cH_{\ell}} \\ = &\; \tr^{(\ell)} \, |S_1
\dots S_{\ell} \, \cU_0^{(\ell)} (s) \, \left( \tr_{\ell+1}  \, [
V_{j,\ell+1},
\gamma^{(\ell+1)}] \right) S_{\ell} \dots S_1| \\
\leq &\;\tr^{(\ell)} \, |S_1 \dots S_{\ell} \, \left( \tr_{\ell+1}
\, V_{j,\ell+1} \, \gamma^{(\ell+1)} \right) S_{\ell} \dots S_1|
\\ &+\tr^{(\ell)} \, |S_1 \dots S_{\ell} \, \left(
\tr_{\ell+1} \, \gamma^{(\ell+1)} \, V_{j,\ell+1} \right) \,
S_{\ell} \dots S_1|
\end{split}
\end{equation}
because the free evolution $\cU_0^{(\ell)}$ commutes with the
operators $S_j$. We consider the first term. By the cyclicity of
the partial trace and because
\[ \tr^{(\ell)} |\tr_{\ell+1} A| \leq \tr^{(\ell+1)} |A|, \]
(see Proposition 9.4 in \cite{EY}), we have
\begin{equation*}
\begin{split}
\tr^{(\ell)} \, |S_1 \dots S_{\ell} &\, \left( \tr_{\ell+1} \,
V_{j,\ell+1} \gamma^{(\ell+1)} \right) S_{\ell} \dots S_1|
\\ &= \tr^{(\ell)} \, |S_1 \dots S_{\ell} \left( \tr_{\ell+1} \,
S_{\ell+1}^{-1} V_{j,\ell+1} \gamma^{(\ell+1)} S_{\ell+1} \right)
S_{\ell} \dots S_1| \\ &\leq \tr^{(\ell+1)} | S_1 \dots S_{\ell}
S_{\ell+1}^{-1} V_{j,\ell+1} \gamma^{(\ell+1)} S_{\ell+1} S_{\ell}
\dots S_1| \\ &= \tr^{(\ell+1)} | \left(S_j S_{\ell+1}^{-1} V_{j,
\ell+1} S_j^{-1} S_{\ell+1}^{-1}\right) S_1 \dots S_{\ell+1}
\gamma^{(\ell+1)} S_{\ell+1} \dots S_1| \\ &\leq \| S_j
S_{\ell+1}^{-1} V_{j,\ell+1} S_j^{-1} S_{\ell+1}^{-1} \| \, \|
\gamma^{(\ell+1)} \|_{\cH_{\ell+1}} \leq \wt C \|\gamma^{(\ell+1)}
\|_{\cH_{\ell+1}}
\end{split}
\end{equation*}
for a constant $\wt C$, independent of $\ell$. In the last
inequality we used Lemma \ref{lm:norm-bound}, part (ii). The
second term on the r.h.s. of (\ref{eq:gammaell}) can be bounded
similarly. We get
\begin{equation*}
\|\cU^{(\ell)} (s) \tr_{\ell+1} [ V_{j,\ell+1} , \gamma^{(\ell+1)}
] \|_{\cH_{\ell}} \leq 2 \wt C \| \gamma^{(\ell+1)}
\|_{\cH_{\ell+1}}\,.
\end{equation*}
Applying this bound iteratively to (\ref{eq:wtzeta}) we find
\begin{equation*}
\| \wt\zeta (k,n,t) \|_{\cH_k} \leq 2^n \lambda^n \wt C^n t^n \,
{k+n \choose n} \, \sup_{s \in [0,t]} \| \gamma^{(k+n)}_{s}
\|_{\cH_{k+n}} \, .
\end{equation*}
By (\ref{eq:apri}) we have, \begin{equation}\label{eq:error} \|
\wt \zeta (k,n,t) \| \leq D^k D^n t^n \, , \end{equation} where
$D$ depends only on the constant $C$ in the bound (\ref{eq:apri})
(and on $\lambda$). Now suppose that $\Gamma_{1,t} = \{
\gamma_{1,t}^{(k)}\}_{k \geq 1}$ and $\Gamma_{2,t} = \{
\gamma^{(k)}_{2,t}\}_{k\geq 1}$ are two solutions of the infinite
BBGKY hierarchy with $\Gamma_{1,t=0} = \Gamma_{2,t=0} = \Gamma_0$
and satisfying (\ref{eq:apri}). Then, for $t \leq 1/(2D)$, we have
\[ \| \gamma^{(k)}_{1,t} - \gamma^{(k)}_{2,t} \|_{\cH_k} = \|\wt
\zeta_1 (k,t,n) - \wt \zeta_2 (k,t,n)\|_{\cH_k} \leq 2 D^k 2^{-n}
\] for any $n>1$ (note that the terms $\zeta (k,t,m)$ in the sum
over $m$ in (\ref{eq:zet}) depend only on the initial data
$\gamma_0^{(k)}$, and thus do not contribute to the difference
$\gamma^{(1)}_{1,t} - \gamma^{(2)}_{2,t}$). Since $n \geq 1$ is
arbitrary, we find
\[ \gamma^{(k)}_{1,t} = \gamma^{(k)}_{2,t} \] for all $k \geq 1$
and for all $t \leq 1/(2D)$. Since the bound (\ref{eq:apri}) holds
uniformly in $t$, for $t\in [0,T]$, the argument can be iterated
to prove that $\Gamma_{1,t} = \Gamma_{2,t}$, for all $t \in
[0,T]$.
\end{proof}

\section{Removal of the cutoffs}
\label{sec:comp}\setcounter{equation}{0}

{F}rom Theorem \ref{thm:convergence} and Theorem \ref{thm:unique},
we know that $\wt \Gamma_{N,t}$ converges to $\Gamma_t = \{
\gamma_t^{(k)} \}_{k\geq 1} \in C([0,T], \cH)$ for $N \to \infty$,
with respect to the product of the topologies $\wh\rho_k$ (defined
in \ref{eq:hatrho}). In this section, we show how to remove the
cutoffs $\eps$ and $\kappa$, which are used to regularize the
interaction and the initial data. To remove the cutoff $\eps$ we
need to compare two dynamics, the one generated by the modified
Hamiltonian (\ref{eq:cutoffham}), and the one generated by the
original Hamiltonian (\ref{eq:ham}): this task is accomplished in
the next proposition.

\begin{proposition}\label{prop:comp} Suppose $\psi_{N,t}^{\eps,\kappa}$ is a
solution of the modified Schr\"odinger equation (\ref{eq:schrcut})
with initial data $\psi_{N}^{\kappa}$, and let
$\psi_{N,t}^{\kappa}$ be the solution of the original
Schr\"odinger equation (\ref{eq:schr}), with the same initial
data. Then there exists a constant $C$, independent of $N$,
$\kappa$, $\eps$ and $t$ such that
\begin{equation*}
\| \psi_{N,t}^{\kappa,\eps} - \psi_{N,t}^{\kappa} \| \leq C \, t
\,  \eps^{1/4}
\end{equation*}
for all $t \geq 0$, and for all $N$ large enough (depending on
$\eps$ and $\kappa$).
\end{proposition}

\begin{proof} We use the notation $\phi_{N,t} = \psi_{N,t}^{\kappa}$
and $\wt \phi_{N,t} = \psi_{N,t}^{\eps,\kappa}$. We compute
\begin{equation*}
\begin{split}
\partial_t \| \phi_{N,t} - \wt \phi_{N,t} \|^2 &= i (H_N \phi_{N,t} -
\wt H_N \wt \phi_{N,t} , \phi_{N,t} - \wt \phi_{N,t}) -i
(\phi_{N,t} - \wt \phi_{N,t}, H_N \phi_{N,t} - \wt H_N \wt \phi_{N,t}) \\
&= - \text{Im} \, (\phi_{N,t} -\wt \phi_{N,t}, (H_N - \wt H_N) \wt
\phi_{N,t})
\end{split}
\end{equation*}
by the self-adjointess of $H_N$. Hence
\begin{equation}\label{eq:comp0}
\pm \partial_t \| \phi_{n,t} - \wt \phi_{N,t} \|^2 \leq \|
\phi_{N,t} - \wt\phi_{N,t}\| \, \|(H_N - \wt H_N) \wt \phi_{N,t}
\| \, .
\end{equation}
We have
\[ H_N - \wt H_N = \frac{\lambda}{N} \sum_{i<j}^N \left( \frac{1}{|x_i
-x_j|} - \frac{1}{|x_i -x_j| +\eps N^{-1}} \right)=
\frac{\lambda}{N} \sum_{i<j}^N \frac{\eps N^{-1}}{|x_i -x_j| (
|x_i -x_j| + \eps N^{-1})}. \] Therefore, using the symmetry with
respect to permutations
\begin{equation}\label{eq:comp}
\begin{split}
\| (H_N - \wt H_N) &\wt \phi_{N,t} \|^2 = (\wt \phi_{N,t} , (H_N -
\wt H_N)^2 \wt \phi_{N,t}) \\ &\leq \lambda^2 \eps^2 \; (\wt
\phi_{N,t}, \frac{1}{|x_1 - x_2| (|x_1 -x_2|+\eps N^{-1})}
\frac{1}{|x_3 - x_4| ( |x_3 - x_4|+\eps N^{-1})} \, \wt \phi_{N,t}) \\
&+ \lambda^2 \eps^2 N^{-1} \; (\wt \phi_{N,t}, \frac{1}{|x_1 -
x_2| (|x_1 -x_2|+\eps N^{-1})} \frac{1}{|x_2 - x_3| ( |x_2 -
x_3|+\eps N^{-1})} \, \wt \phi_{N,t})\\ &+ \lambda^2 \eps^2 N^{-2}
\; (\wt \phi_{N,t}, \frac{1}{|x_1 - x_2|^2 (|x_1 -x_2|+\eps
N^{-1})^2} \, \wt \phi_{N,t}) \, .
\end{split}
\end{equation}
The first term on the r.h.s. of the last equation can be estimated
by
\begin{equation*}
\begin{split}
(\wt \phi_{N,t}, \frac{1}{|x_1 - x_2| (|x_1 -x_2|+\eps N^{-1})}
&\frac{1}{|x_3 - x_4| ( |x_3 - x_4|+\eps N^{-1})} \, \wt
\phi_{N,t})
\\ &\hspace{1cm}\leq
(\wt \phi_{N,t}, \frac{1}{|x_1-x_2|^2 \, |x_3 - x_4|^2} \, \wt \phi_{N,t}) \\
&\hspace{1cm} \leq C \, (\wt \phi_{N,t}, \, S_1^2 S_2^2 S_3^2
S_4^2 \, \wt \phi_{N,t}) \leq C
\end{split}
\end{equation*}
for all $t \in \bR$, and for every $N$ large enough (depending on
$\eps,\kappa$, see Theorem \ref{thm:apriori}). The second term can
be bounded, using again Theorem \ref{thm:apriori}, by
\begin{equation*}
\begin{split}
(\wt \phi_{N,t}, \frac{1}{|x_1 - x_2| (|x_1 -x_2|+\eps N^{-1})}
&\frac{1}{|x_2 - x_3| ( |x_2 - x_3|+\eps N^{-1})} \, \wt
\phi_{N,t}) \\ &\leq \eps^{-1} N (\wt \phi_{N,t},
\frac{1}{|x_1-x_2|^2 \, |x_2 - x_3|} \, \wt \phi_{N,t}) \\ &\leq C
\, \eps^{-1} N (\wt
\phi_{N,t}, S_3 \frac{1}{|x_1 -x_2|^2} S_3 \, \wt \phi_{N,t}) \\
&\leq C \, \eps^{-1} N (\wt \phi_{N,t}, \, S_1^2 S_2^2 S_3^2 \,
\wt \phi_{N,t})
\\ &\leq C \, \eps^{-1} N
\end{split}
\end{equation*}
for all $N$ large enough. Finally, we estimate the last term on
the r.h.s. of (\ref{eq:comp}) as follows
\begin{equation}\label{eq:comp2}
\begin{split}
(\wt \phi_{N,t}, \frac{1}{|x_1 - x_2|^2 (|x_1 -x_2|+\eps
N^{-1})^2} \, \wt \phi_{N,t}) &\leq \eps^{-3/2}
N^{3/2} (\wt \phi_{N,t}, \frac{1}{|x_1-x_2|^{5/2}} \, \wt \phi_{N,t}) \\
&\leq C \eps^{-3/2} N^{3/2} (\wt \phi_{N,t}, \, S_1^{5/2}
S_2^{5/2} \, \wt \phi_{N,t})
\end{split}
\end{equation}
by Lemma \ref{lm:norm-bound}, part (i). Next we note that, by
Theorem \ref{thm:apriori},
\[ (\wt \phi_{N,t}, S_1^3 S_2^3 \wt \phi_{N,t}) \leq (\wt \phi_{N,t}, S_1^4
S_2^2 \wt \phi_{N,t}) \leq C N \] for all $N$ large enough, and
for all $t$ (the constant $C$ is independent of $\eps$). Since
$(S_1 S_2)^{1/2} \leq N^{1/2} + N^{-1/2} S_1 S_2$, we find
\[ (\wt \phi_{N,t}, S_1^{5/2} S_2^{5/2} \wt \phi_{N,t}) \leq N^{1/2}
(\wt \phi_{N,t}, S_1^2 S_2^2 \phi_{N,t}) + N^{-1/2} (\wt
\phi_{N,t}, S_1^3 S_2^3 \wt \phi_{N,t}) \leq C N^{1/2}\,. \]
{F}rom (\ref{eq:comp2}) we get
\[ (\wt \phi_{N,t}, \frac{1}{|x_1 - x_2|^2 (|x_1 -x_2|+\eps N^{-1})^2}
\, \wt \phi_{N,t}) \leq C \eps^{-3/2} N^2 \, \] for all $N$ large
enough. {F}rom (\ref{eq:comp}) it follows that \[ \| (H_N - \wt
H_N) \wt \phi_{N,t} \|^2 \leq C \eps^{1/2} \] for all $N$ large
enough (depending on $\eps$), and for all $t$ (the constant $C$
only depends on $\lambda$ and $\| \ph\|_{H^1}$, see Theorem
\ref{thm:apriori}). Hence, by (\ref{eq:comp0}) and by Gronwall's
Lemma,
\begin{equation*}
\pm \partial_t \| \phi_{N,t} - \wt \phi_{N,t} \|^2 \leq C
\eps^{1/4} \| \phi_{N,t} - \wt \phi_{N,t} \| \quad \Rightarrow
\quad \| \phi_{N,t} - \wt \phi_{N,t} \| \leq C \eps^{1/4} t
\end{equation*}
for all $t \geq 0$ and for all $N$ large enough.
\end{proof}

Finally we have to remove the cutoff $\kappa$ from the initial
wave function $\psi_{N}^{\kappa}$.

\begin{proposition}\label{prop:kappa}
Suppose $\ph \in H^1 (\bR^3)$, with $\| \ph \| = 1$. For $\kappa
>0$ put $\ph^{\kappa} = \exp (-\kappa |p|/N) \ph$. Suppose
$\psi_{N,t}^{\kappa}$ and $\psi_{N,t}$ are the solutions of the
Schr\"odinger equation (\ref{eq:schr}) with initial data
$\psi_N^{\kappa} (\bx) = \prod_{j=1}^N \ph^{\kappa} (x_j)$, and,
respectively, $\psi_{N}(\bx) = \prod_{j=1}^k \ph (x_j)$. Then
\begin{equation*}
\| \psi_{N,t}^{\kappa} - \psi_{N,t}\| \leq \kappa \| \ph \|_{H^1}
\end{equation*}
for all $t\in \bR$ and all $N\geq 1$.
\end{proposition}

\begin{proof}
By the unitarity of the time evolution, we have \begin{equation}
\begin{split}\label{eq:kappa1}
\| \psi_{N,t}^{\kappa} - \psi_{N,t} \| &= \left\| \psi_N^{\kappa}
- \psi_N \right\| = \left\| \prod_{j=1}^N \ph^{\kappa} (x_j)
-\prod_{j=1}^N \ph (x_j) \right\| \\ &\leq \sum_{j=1}^N \left\|
\ph^{\kappa} (x_1) \dots \ph^{\kappa} (x_{j-1}) (\ph^{\kappa}
(x_j) - \ph (x_j)) \ph (x_{j+1}) \dots \ph (x_N) \right\| \\ &\leq
N \| \ph^{\kappa} -\ph\|\,.
\end{split}
\end{equation}
Here we used that the $L^2$ norm of $\ph^{\kappa}$ is bounded by
$\| \ph^{\kappa} \| \leq \| \ph \| =1$. Since $\ph^{\kappa} =
e^{-\kappa |p|/N} \ph$ and $1- e^{-\kappa|p|/N} \leq \kappa N^{-1}
|p|$, we have
\begin{equation*}
\begin{split}
\| \ph^{\kappa} - \ph \|^2 = \int \rd p \, \left( e^{-\kappa|p|/N}
-1 \right)^2 |\widehat \ph (p)|^2 \leq \kappa^2 N^{-2} \int \rd p
\, |p|^2  |\widehat \ph (p)|^2 = \kappa^2 N^{-2} \| \ph \|_{H^1}^2
\,.
\end{split}
\end{equation*}
By (\ref{eq:kappa1}), we find
\begin{equation*}
\| \psi_{N,t}^{\kappa} - \psi_{N,t} \| \leq \kappa \| \ph
\|_{H^1}\,.
\end{equation*}
\end{proof}

\section{Some Technical Results}
\label{sec:techn}\setcounter{equation}{0}

In this section we collect technical results used throughout the
paper.

In the first lemma we show how to control singularities like $|x_1
- x_2|^{-a}$, for $a<3$ in terms of the operators $S_j =
(1+p_j^2)^{1/4} = (1-\Delta_j)^{1/4}$, for $j=1,2$.

\begin{lemma}\label{lm:norm-bound}
Let $V(x) = |x|^{-1}$ and $V^{\eps} (x) = (|x| + \eps
N^{-1})^{-1}$. Moreover, we set $V_{i,j} = V(x_i -x_j)$ and
$V^{\eps}_{i,j} = V^{\eps} (x_i-x_j)$.
\begin{itemize}
\item[i)] For all $a<3$ there exists $C(a)$ such that
\begin{equation*} \frac{1}{|x_1 -x_2|^a} \leq C(a)
\, S_1^{\alpha} S_2^{\beta} \quad \quad \text{for all } \alpha,
\beta >0 \quad \text{with } \alpha+ \beta =2a \, .
\end{equation*} If $a=1$, we have the tight bound
\begin{equation*}
V_{1,2} \leq \frac{\pi}{2} \, S_1^{\alpha} S_2^{\beta} \quad \quad
\text{for all } \alpha, \beta >0 \quad \text{with } \alpha+ \beta
=2 \, .\end{equation*}  \item[ii)] The operator \[ S_1 S_2^{-1}
V_{1,2} S_1^{-1} S_2^{-1}
\] is bounded. Moreover \begin{equation}\label{eq:boueps}
\| S_1 S_2^{-1} V^{\eps}_{1,2} S_1^{-1} S_2^{-1} \| \leq C
\end{equation}
uniformly in $\eps$. \item[iii)] For all $\delta >0$,
there exists $C_{\delta} <\infty$ such that
\begin{equation*}
\| S_1^{-2} S_2^{-1} V^{\eps}_{1,2} S_2 \| \leq C_{\delta}
\eps^{-\delta} N^{\delta}
\end{equation*}
\end{itemize}
\end{lemma}
\begin{proof}
To prove i) we use that
\begin{equation}\label{eq:herbst}
\frac{1}{|x_1 - x_2|} \leq \frac{\pi}{2} S_i^2, \quad \text{and }
\quad \frac{1}{|x_1 - x_2|^a} \leq C(a) S_i^{2a}
\end{equation}
for some constant $C(a) < \infty$, and for $i=1,2$. This is proven
in \cite{He}. The statement i) can now be shown by the following
general result. Suppose $A \geq 0$, and $B,C \geq 1$ are three
operators, with \[ A \leq B \quad \quad \text{and } \quad \quad A
\leq C \] and such that $B$ commutes with $C$. Then
\begin{equation}\label{eq:matrix}
A \leq B^{\alpha} C^{\beta} \quad \text{for all } \alpha,\beta
\quad \text{with } \alpha + \beta =1 \, .\end{equation} In fact,
$A \leq B$ implies that $B^{-1/2} A B^{-1/2} \leq 1$, and, by the
operator monotonicity of powers smaller than one, this implies
that $(B^{-1/2} A B^{-1/2})^{\alpha} \leq 1$ for all $\alpha \leq
1$. On the other hand $A \leq C$ implies $B^{-1/2} A B^{-1/2} \leq
B^{-1} C$ and also $(B^{-1/2} A B^{-1/2})^{\beta} \leq B^{-\beta}
C^{\beta}$ for all $\beta\leq 1$. Hence
\begin{equation*}\begin{split} B^{-1/2} A B^{-1/2} &= (B^{-1/2} A
B^{-1/2})^{\beta/2} (B^{-1/2} A B^{-1/2})^{\alpha} (B^{-1/2} A
B^{-1/2})^{\beta/2} \\ &\leq (B^{-1/2} A B^{-1/2})^{\beta} \leq
B^{-\beta} C^{\beta} \,.\end{split}\end{equation*} Multiplying
both sides by $B^{1/2}$ gives (\ref{eq:matrix}). Part i) follows
by (\ref{eq:herbst}) and by (\ref{eq:matrix}), taking $A=|x_1
-x_2|^{-a}$, $B=S_1$ and $C=S_2$. Next we prove ii). We will make
use of following fact (known as the Holmgren-Shur inequality).
Suppose that the set of mutually orthognal projections
$\{P_n\}_{n=1}^\infty$ resolves the identity in a strong  sense,
that is
\[s-\lim_{N\rightarrow\infty}\sum_{n=1}^N P_n=1\,.\] Then for any
operator $A$ we have the norm inequality
\begin{equation}\label{eq:prop}
\|A\| \ \le \max{\left(\sup_k \sum_{n=1}^{\infty} \|P_k\, A\,
P_n\|\,,\,\sup_n \sum_{k=1}^{\infty} \|P_k\, A\, P_n\|\right)}\,.
\end{equation}
We choose
\[P_k:=\chi (16^{k-1} \le p_1^2+1 < 16^k)\,,\] then the boundedness of
$S_1 S_2^{-1} V_{1,2} S_1^{-1} S_2^{-1}$ follows by
(\ref{eq:prop}) if we prove that
\begin{equation}\label{eq:bnd1}
\|P_k\,S_1^{-1}\,S_2^{-1}\,\frac{1}{|x_1-x_2|}\,S_1\,S_2^{-1}\,P_n\|
\ \le C \; 2^{-|k-n|/2}\ \,.
\end{equation}
for a constant $C$ independent of $n,k$. To show the estimate
(\ref{eq:bnd1}) we consider two possibilities: $k\ge n$ and $n>k$.

When $k\ge n$ we can bound the l.h.s. of (\ref{eq:bnd1}) by
\begin{equation}\label{eq:bnd2}
\|P_k\,S_1^{-1}\|\,\|S_2^{-1}\,\frac{1}{|x_1-x_2|}\,S_2^{-1}\|\,\|S_1\,P_n\|
\,.
\end{equation}
Since, by part i), \[
\|\,S_2^{-1}\,\frac{1}{|x_1-x_2|}\,S_2^{-1}\| \ \leq \
\frac{\pi}{2}\,,\] and \[\|P_k\,S_1^{-1}\| \le \ 2^{-k+1}\,,\quad
\|S_1\,P_n\| \ \le \ 2^{n}\,,\] we obtain (\ref{eq:bnd1}) (in fact
a stronger estimate). If $n>k$ we write
\begin{multline}\label{eq:bnd3}
P_k\,S_1^{-1}\,S_2^{-1}\,\frac{1}{|x_1-x_2|}\,S_1\,S_2^{-1}\,P_n
\\ = P_k\,S_1^{-1}\,S_2^{-1}\,(p_1^2+1)(p_1^2+1)^{-1}\,
\frac{1}{|x_1-x_2|}\,S_1\,S_2^{-1}\,P_n\\
=P_k\,S_1^{-1}\,S_2^{-1}\,(p_1^2+1)\,\frac{1}{|x_1-x_2|}\,
(p_1^2+1)^{-1}\,S_1\,S_2^{-1}\,P_n
\\ +P_k\,S_1^{-1}\,S_2^{-1}\,(p_1^2+1)\,
\com{(p_1^2+1)^{-1}}{\frac{1}{|x_1-x_2|}}\,S_1\,S_2^{-1}\,P_n
\,.
\end{multline}
The norm of the first contribution can be now bounded by
\[\|P_k\,S_1^{3}\|\,\|S_2^{-1}\,\frac{1}{|x_1-x_2|}\,S_2^{-1}\|\,\|S_1^{-3}\,P_n\|
\ \le \  C \; 2^{-3\,|k-n|}\,,\] hence we are left with the task
of checking that also the norm of second contribution in
(\ref{eq:bnd3}) satisfies a similar bound. Note now that
\[\com{(p_1^2+1)^{-1}}{\frac{1}{|x_1-x_2|}} \ = \
(p_1^2+1)^{-1}\,\left\{p_1\cdot\frac{x_1-x_2}{|x_1-x_2|^3} \ + \
\frac{x_1-x_2 }{|x_1-x_2|^3}\cdot p_1
\right\}\,(p_1^2+1)^{-1}\,,\]therefore it suffices to estimate
\begin{equation}\label{eq:bnd4}
\|P_k\,S_1^{-1}\,S_2^{-1}\,
p_1\cdot\frac{x_1-x_2}{|x_1-x_2|^3}\,(p_1^2+1)^{-1}\,\,S_1\,S_2^{-1}\,P_n\|
\end{equation}
and
\begin{equation}\label{eq:bnd5}
\|P_k\,S_1^{-1}\,S_2^{-1}\,\frac{x_1-x_2}{|x_1-x_2|^3}\cdot
p_1\,(p_1^2+1)^{-1}\,\,S_1\,S_2^{-1}\,P_n\| \,.
\end{equation}
We bound (\ref{eq:bnd4}), using the result of part i), by
\[\|P_k\,S_1^{2}\|\,\|S_1^{-3}\,S_2^{-1}\,p_1\cdot\frac{x_1-x_2}{|x_1-x_2|^3}
\,S_2^{-1}S_1^{-1}\|\,\|S_1^{-2}\,P_n\| \ \le \ C \; 2^{-2\,
|k-n|}\,.\] As for (\ref{eq:bnd5}) we estimate it by
\begin{equation*}
\begin{split}
\|P_k\,S_1^{1/2}\|\,\|S_1^{-3/2}&\,S_2^{-1}\,\frac{x_1-x_2
}{|x_1-x_2|^3}\cdot
p_1\,S_2^{-1}\,S_1^{-5/2}\|\,\|S_1^{-1/2}\,P_n\| \\ &\leq   C \;
2^{-|k-n|/2} \, \| S_1^{-3/2} S_2^{-1} \frac{1}{|x_1 -x_2|^{5/4}}
\|
\, \| \frac{1}{|x_1-x_2|^{3/4}} S_1^{-1/2} S_2^{-1} \| \\
&\leq C \, 2^{-|k-n|/2} \,,
\end{split}
\end{equation*} hence the
result. The same line of reasoning can clearly be applied also to
prove (\ref{eq:boueps}), uniformly in $\eps>0$. Finally, we show
part iii). To this end, we write
\begin{equation*}
S_1^{-2} S_2^{-1} V^{\eps}_{1,2} S_2 = S_1^{-2} V^{\eps}_{1,2} +
\int_0^{\infty} \rd s \, s^{-1/4} \, \frac{S_1^{-2}}{s+1 +p_2^2}
\left(p_2 \cdot \nabla V^{\eps}_{1,2} + \nabla V^{\eps}_{1,2}
\cdot p_2 \right) \frac{S_2}{s+1+p_2^2} \, .
\end{equation*}
The first term is bounded, by part i), uniformly in $\eps$. As for
the integral, for any given $\delta >0$ its norm can be bounded by
\begin{equation*}
\begin{split}
(\eps^{-1} N)^{\delta} \, \int_0^{\infty} &\frac{\rd s}{ s^{1/4}}
\left\{ \left\| \frac{|p_2|}{s+1 +p_2^2} \right\| \, \left\|
S_1^{-2} \frac{1}{|x_1-x_2|}\right\| \left\| \frac{1}{|x_1
-x_2|^{1-\delta}} \, \frac{1}{(|p_2| + 1)^{1-\delta}} \right\|
\left\| \frac{S_2 (|p_2|+1)^{1-\delta}}{s+1 + p_2^2} \right\| \right. \\
&\hspace{2cm} + \left\| \frac{(|p_2| + 1)^{1/2 -
\delta}}{s+1+p_2^2} \right\| \, \left\| (|p_2| +1)^{-1/2 + \delta}
S_1^{-2} \frac{1}{|x_1-x_2|^{3/2 - \delta}}\right\| \\
&\left.\hspace{4.5cm} \times \left\| \frac{1}{|x_1 -x_2|^{1/2}} \,
\frac{1}{(|p_2| + 1)^{1/2}} \right\|
\left\| \frac{S_2 |p_2| (|p_2|+1)^{1/2}}{s+1 + p_2^2} \right\| \right\} \\
&\leq D_{\delta} \eps^{-\delta} N^{\delta} \int_0^{\infty} \rd s
\, s^{-1/4} \frac{1}{(s+1)^{3/4 + \delta/2}} \\ &\leq C_{\delta}
\eps^{-\delta} N^{\delta} \,.
\end{split}
\end{equation*}
\end{proof}

The following lemma is used in the proof of Proposition
\ref{prop:upper}, in order to control terms of the form
$V_{1,2}^{\eps} S_1^{2n} V_{1,2}^{\eps}$ by powers of $S_1$ and
$S_2$.

\begin{lemma}\label{lm:comm}
For every $n \geq 1$, $\eps >0$, and $0<a<1$, there exists $D
(n,\eps,a)$, independent of $N$, such that
\begin{equation*}
\begin{split}
V_{1,2}^{\eps} S_1^{2n} V_{1,2}^{\eps} \leq \; &D(n,\eps,a) \, N^a
\,  \sum_{m=1}^{n} S_1^{2(n-m+2)} S_2^2 N^{m-1}
\end{split}
\end{equation*}
\end{lemma}
\begin{proof}
Suppose first that $n = 2\ell$ is even. Then
\[ S_1^{2n} = S_1^{4\ell} = (1 + p_1^2)^{\ell} \leq 2^{\ell} \, (1 + (p_1^2)^{\ell}) \leq 2^{\ell} +
6^{\ell} \sum_{\alpha = 1}^3 p_{1,\alpha}^{2\ell}\, ,\] where
$p_{1,\alpha}$ denotes the $\alpha$'s component of the vector
$p_1$. Hence
\begin{equation}\label{eq:lem1}
\begin{split}
V_{1,2}^{\eps}\, S_1^{4\ell} \,V_{1,2}^{\eps} &\leq 2^{\ell} \,
(V_{1,2}^{\eps})^2 + 6^{\ell} \sum_{\alpha=1}^3 V_{1,2}^{\eps} \,
p_{1,\alpha}^{2\ell} \, V_{1,2}^{\eps} \, .
\end{split}
\end{equation}
Using the commutator expansion \[ V_{1,2}^{\eps} \,
p_{1,\alpha}^{\ell} = p_{1,\alpha}^{\ell} \, V_{1,2}^{\eps} +
\sum_{j=1}^{\ell} {\ell \choose j} \, p_{1,\alpha}^{\ell-j} \,
\underbrace{[ \dots [ V_{1,2}^{\eps},
p_{1,\alpha}],p_{1,\alpha}],\dots p_{1,\alpha}]}_{\text{$j$
commutators}}
\] and a Schwarz inequality, we obtain the bound
\begin{equation}\label{eq:VSV}
\begin{split}
V_{1,2}^{\eps} \, p_{1,\alpha}^{2\ell} \, V_{1,2}^{\eps} \leq 2\,
p_{1,\alpha}^{\ell} \, (V_{1,2}^{\eps})^2 \, p_{1,\alpha}^{\ell} +
2^{\ell} \sum_{j=1}^{\ell} {\ell \choose j} p_{1,\alpha}^{\ell-j}
\; |\underbrace{[ \dots [V_{1,2}^{\eps}, p_{1,\alpha}], \dots
p_{1,\alpha}] \dots ]}_{\text{$j$ commutators}}|^2 \,
p_{1,\alpha}^{\ell-j} \, .
\end{split}
\end{equation}
Next we use $(V_{1,2}^{\eps})^2 \leq C \, S_1^2 S_2^2$ and
\begin{equation*}\begin{split} |\underbrace{[ \dots [V_{1,2}^{\eps},
p_{1,\alpha}], \dots p_{1,\alpha}] \dots ]}_{\text{$j$
commutators}}|^2 &\leq C (j!)^2 \, \frac{1}{(|x_1 - x_2| +\eps
N^{-1})^{2j+2}} \\& \leq C (j!)^2 (\eps^{-1} N)^{2j-1+a}
\frac{1}{|x_1-x_2|^{3-a}} \\ &\leq D_1 (j,\eps,a) \, N^{2j-1+a} \,
S_1^4 S_2^2 \end{split}
\end{equation*} for some $0< a < 1$ (the constant
$D_1 (j,\eps,a)$ is proportional to $(j!)^2$, to $\eps^{-2j+1-a}$,
and diverges logarithmically in $a$, for $a \to 0$). {F}rom
(\ref{eq:VSV}) we find
\begin{equation*}
\begin{split}
V_{1,2}^{\eps} \, p_{1,\alpha}^{2\ell} \, V_{1,2}^{\eps} \leq C\,
p_{1,\alpha}^{\ell} \, S_1^2 S_2^2  \, p_{1,\alpha}^{\ell} +
D_2(\ell,\eps,a) \, \sum_{j=1}^{\ell} p_{1,\alpha}^{2(\ell-j)}
S_1^4 S_2^2 \, N^{2j-1+a}
\end{split}
\end{equation*}
for a new constant $D_2 (\ell,\eps,a)$ independent of $N$.
Inserting the last equation into (\ref{eq:lem1}), summing over
$\alpha$, and using again $(V_{1,2}^{\eps})^2 \leq C S_1^2 S_2^2$,
it follows that
\begin{equation*}
V_{1,2}^{\eps} S_1^{4\ell} V_{1,2}^{\eps} \leq D(\ell,\eps,a) \,
\left( S_1^{4\ell + 2} S_2^2 + \sum_{j=1}^{\ell} S_1^{4(\ell-j)}
S_1^4 S_2^2 \, N^{2j-1+a} \right)\,.
\end{equation*}
Replacing the indices $\ell$ and $j$ by $n=2\ell$ and $m=2j$, we
have
\begin{equation*}
V_{1,2}^{\eps} S_1^{2n} V_{1,2}^{\eps} \leq C(n,\eps,a)
\sum_{m=1}^n S_1^{2(n-m+2)} S_2^2 \, N^{m-1+a}
\end{equation*}
which proves the claim if $n$ is even (since all terms in the sum
over $m$ are positive, we can also allow $m$ to be odd).

If $n$ is odd, the proof is a little bit more difficult. Let
$n=2\ell+1$. Then
\begin{equation}\label{eq:lem3}
\begin{split}
V_{1,2}^{\eps} S_1^{4\ell+2} V_{1,2}^{\eps} = \; &V_{1,2}^{\eps}
(1+p_1^2)^{\ell} S_1^{2} V_{1,2}^{\eps} \\ \leq \; &2^{\ell}
V_{1,2}^{\eps} S_1^2 V_{1,2}^{\eps} + 6^{\ell} \sum_{\alpha=1}^3
V_{1,2}^{\eps} p_{1,\alpha}^{2\ell} S_1^2 V_{1,2}^{\eps} \\ \leq
\; &C (\ell) \left( S_1 (V_{1,2}^{\eps})^2 S_1 + |[V_{1,2}^{\eps},
S_1]|^2+ \sum_{\alpha=1}^3 p_{1,\alpha}^{\ell}  S_1
(V_{1,2}^{\eps})^2 S_1 p_{1,\alpha}^{\ell} + p_{1,\alpha}^{\ell}
|[ V_{1,2}^{\eps}, S_1]|^2 p_{1,\alpha}^{\ell} \right. \\
&\hspace{2cm} + \sum_{\alpha=1}^3 \sum_{j=1}^{\ell} \,
p_{1,\alpha}^{\ell-j} S_1 \, | \underbrace{[ \dots [
V_{1,2}^{\eps}, p_{1,\alpha}], \dots ] ,p_{1,\alpha}]}_{\text{$j$
commutators}} |^2 S_1 p_{1,\alpha}^{\ell-j} \\ &\hspace{2cm}
\left. + \sum_{\alpha=1}^3 \sum_{j=1}^{\ell} \,
p_{1,\alpha}^{\ell-j} \; | [\underbrace{[ \dots [ V_{1,2}^{\eps},
p_{1,\alpha}] \dots ] ,p_{1,\alpha}]}_{\text{$j$ commutators}} ,
S_1] |^2 p_{1,\alpha}^{\ell-j} \right)\,.
\end{split}
\end{equation}
Next we note that, for any integer $r \geq 0$,
\begin{equation*}
\left[ \frac{\partial^{r}}{\partial x^r_{1,\alpha}} V_{1,2}^{\eps}
, S_1 \right] = \sum_{\beta=1}^3 \int_0^{\infty} \rd s \, s^{1/4}
\frac{1}{s+1+p_1^2} \, \left( p_{1,\beta}
\frac{\partial^{r+1}}{\partial x_{1,\beta} \partial
x^r_{1,\alpha}} V_{1,2}^{\eps} + \text{h.c.} \right) \,
\frac{1}{s+1 + p_1^2}\,.
\end{equation*}
Using \[ \left| \frac{\partial^{r+1} V_{1,2}^{\eps}}{\partial
x_{1,\beta} \partial x_{1,\alpha}^r} \right| \leq C  \,
\frac{(r+1)!}{(|x_1-x_2|+\eps N^{-1})^{r+2}}, \] we find, for
every $r \geq 0$, $\eps >0$ and $a >0$, a constant $C(r,\eps,a)$,
such that
\begin{equation*}
\left\| \left[ \frac{\partial^{r}}{\partial x^r_{1,\alpha}}
V_{1,2}^{\eps} , S_1 \right] S_1^{-2} S_2^{-1} \right\|  \leq
C(r,\eps,a) N^{r+(a/2)}
\end{equation*}
and thus, for every $a>0$, we have the operator inequality
\begin{equation*}
\left|\left[ \frac{\partial^{r}}{\partial x^r_{1,\alpha}}
V_{1,2}^{\eps} , S_1 \right]\right|^2 \leq C(r,\eps,a) S_1^4 S_2^2
\, N^{2r + a}\,.
\end{equation*}
This can be used in the second, fourth and last term on the r.h.s.
of (\ref{eq:lem3}). The other terms can be handled as in the case
of even $n$.
\end{proof}

In the next lemma we give a proof of the fact that a solution
$\ph_t$ of the Hartree equation (\ref{eq:hartree2}) has
$H^{1/2}$-norm uniformly bounded in time. We assume here that the
initial data $\ph \in H^{1/2}$ (we apply this result in the proof
of Theorem \ref{thm:main}, Step 5, where we have the stronger
condition $\ph \in H^1 (\bR^3)$).

\begin{lemma}\label{lm:hartree}
Suppose $\ph \in H^{1/2} (\bR^3)$, with $\| \ph \| =1$, and let
$\ph_t$ be the solution of the nonlinear Hartree equation
\[ i\partial_t \ph_t = (1-\Delta)^{1/2} \ph_t + \lambda
(\frac{1}{|\, . \, |} *|\ph_t|^2) \ph_t \] with initial data
$\ph_{t=0} = \ph$. Then, if $\lambda > -4/\pi$, there exists a
constant $C$, depending only on $\lambda$ such that
\[ (\ph_t , (1- \Delta)^{1/2} \ph_t) \leq C \| \ph \|_{H^{1/2}} \]
for all $t\in \bR$.
\end{lemma}
\begin{proof}
The $L^2$ norm of $\ph$ is conserved, so $\| \ph_t \| =1$. Also
the Hartree energy
\[ E(\ph) = \int \rd x \, |(1-\Delta)^{1/4} \ph (x)|^2 +
\frac{\lambda}{2} \int \rd x \rd y \, \frac{1}{|x-y|} \, |\ph
(x)|^2 |\ph (y)|^2 \] is conserved by the time evolution. Note
that
\begin{equation*}
\begin{split}
\int \rd x \rd y \, \frac{1}{|x-y|} \, |\ph (x)|^2 |\ph (y)|^2
&\leq \sup_{x} \int \rd y \,  \frac{1}{|x-y|} |\ph (y)|^2 \, \|
\ph \|^2
\\ &\leq \frac{\pi \| \ph \|^2}{2} \, \int \rd y \, |(1-\Delta_y)^{1/4} \ph (y)|^2
\end{split}\end{equation*}
where we used the operator inequality $|x-y|^{-1} \leq (\pi/2) (1
- \Delta_y)^{1/2}$ for every $x \in \bR^3$. For $\lambda <0$, we
find
\[ (1 + \frac{\pi}{4} \lambda) (\ph, (1- \Delta)^{1/2} \ph) \leq E(\ph) \leq
(\ph,(1-\Delta)^{1/2} \ph) \,.\] For $\lambda >0$, on the other
hand, we have
\[ (\ph, (1-\Delta)^{1/2} \ph) \leq E(\ph) \leq
(1 + \frac{\pi}{4} \lambda) (\ph, (1-\Delta)^{1/2} \ph) \, .\]
Hence, for all $\lambda > - 4/\pi$, we have
\[ (\ph_t, (1-\Delta)^{1/2} \ph_t) \leq C E(\ph_t) = C E(\ph) \leq C \| \ph
\|^2_{H^{1/2}} \] for a constant $C$ only depending on $\lambda$.
\end{proof}

Finally, in the next lemma, we give a criterium for the
equicontinuity of a sequence of time dependent density matrices
$\gamma_{N,t}^{(k)} \in \cH_k$, with respect to the metric
$\rho_k$. This result is used in the proof of Theorem
\ref{thm:compact}, in Step 3, to show the compactness of the
sequence $\wt \gamma_{N,t}^{(k)} \in C([0,T], \cH_k)$ with respect
to the metric $\wh \rho_k$.

\begin{lemma}\label{lm:equicont}
A sequence of time-dependent density matrices $\gamma_{N,
t}^{(k)}$, $N=1,2, \ldots$, defined for  $t \in [0,T]$ and
satisfying \begin{equation}\label{eq:rec} \sup_{t\in [0,T]} \|
\gamma_{N,t}^{(k)} \|_{\cH_k} \leq C\end{equation} for all $N$, is
equicontinuous in $C([0,T], \cH_k)$ with respect to the metric
$\rho_k$ (defined in (\ref{eq:rho})), if and only if, for all
$J^{(k)}$ in a dense subset of $\cA_k$, and for every $\eta
>0$ there exists a $\delta
> 0$ such that
\begin{equation}\label{eq:equi02}
\Big| \tr \;  J^{(k)} \left( \gamma_{N,t}^{(k)} -
\gamma_{N,s}^{(k)} \right) \Big| \leq \eta
\end{equation}
for all $N$, whenever $|t -s| \leq \delta$.
\end{lemma}

\begin{proof}
Equicontinuity w.r.t. the  metric $\rho_k$ means that, for any
$\eta
> 0$ there exists $\delta
>0$ (independent of $N$), such that
\begin{equation}\label{eq:equi0}
\rho_k ( \gamma^{(k)}_{N,t} , \gamma^{(k)}_{N,s} ) = \sum_{j
=1}^{\infty} 2^{-j} \left| \tr \;  J_j^{(k)}
\left(\gamma_{N,t}^{(k)} - \gamma_{N,s}^{(k)} \right) \right| \leq
\eps
\end{equation}
if $|t -s| \leq \delta$. Recall that $\{ J^{(k)}_j \}_{j \geq 1}$
was chosen as a dense countable subset of the unit ball of
$\cA_k$. Using (\ref{eq:rec}), one can approximate any $J^{(k)}
\in \cA_k$ by an appropriate finite linear combinations of the
$J^{(k)}_j$ and thus one can easily prove that (\ref{eq:equi0})
implies (\ref{eq:equi02}).

Next we prove the opposite implication. By a standard
approximation argument, one can prove that, if (\ref{eq:equi02})
holds for all $J^{(k)}$ in a dense subset of $\cA_k$, then it
holds for every $J^{(k)}$ in $\cA_k$. In particular, for every $j
\geq 1$, and $\eta >0$, we can find $\delta (j,\eta)
>0$ such that
\begin{equation*}
\Big| \tr \;  J_j^{(k)} \left( \gamma_{N,t}^{(k)} -
\gamma_{N,s}^{(k)} \right) \Big| \leq \eta
\end{equation*}
for all $N$, if $|t -s| \leq \delta (j,\eta)$. Moreover, using
(\ref{eq:rec}), we note that, given $\eta >0$, we have
\begin{equation*}
\sum_{j > m} 2^{-j} \Big|\tr \; J_j^{(k)} \left(
\gamma_{N,t}^{(k)} - \gamma_{N,s}^{(k)} \right) \Big| \leq \sum_{j
>m} 2^{-j} \| J^{(k)}_j \|_{\cA_k} \left( \| \gamma^{(k)}_{N,t} \|_{\cH_k} +
\|\gamma^{(k)}_{N,s} \|_{\cH_k} \right) \leq C \sum_{j >m} 2^{-j}
\leq \eta/2
\end{equation*}
if $m$ is sufficiently large (independently of $N$ and of $t,s \in
[0,T]$). Hence
\begin{equation*}
\sum_{j \geq 1} 2^{-j} \Big| \tr \; J_j^{(k)} \left(
\gamma_{N,t}^{(k)} - \gamma_{N,s}^{(k)} \right) \Big| \leq \eta/2
+ \sum_{j \leq m} 2^{-j} \Big| \tr \; J_j^{(k)} \left(
\gamma_{N,t}^{(k)} - \gamma_{N,s}^{(k)} \right) \Big|.
\end{equation*}
With $\delta = \min_{j \leq m} \delta (j,\eta/2)$, we find
\begin{equation*}
\sum_{j \geq 1} 2^{-j} \Big| \tr \; J_j^{(k)} \left(
\gamma_{N,t}^{(k)} - \gamma_{N,s}^{(k)} \right) \Big| \leq \eta
\end{equation*}
for all $t,s \in [0,T]$ with $|t-s| \leq \delta$ and for all $N$.
This proves that (\ref{eq:equi02}) implies (\ref{eq:equi0}).
\end{proof}


\thebibliography{hh}

\bibitem{BGM}
C. Bardos, F. Golse and N. Mauser: {\sl Weak coupling limit of the
$N$-particle Schr\"odinger equation.} Methods Appl. Anal. {\bf 7}
(2000) 275--293.

\bibitem{EESY}
A. Elgart, L. Erd\H os, B. Schlein, and H.-T. Yau: {\sl
Gross-Pitaevskii Equation as the Mean Field Limit of Weakly
Coupled Bosons.} arXiv:math-ph/0410038. To appear in Arch. Rat.
Mech. Anal.

\bibitem{ESY} L. Erd{\H{o}}s, B. Schlein and H.-T. Yau: {\sl Derivation
of the {G}ross-{P}itaevskii Equation for the Dynamics of
{B}ose-{E}instein Condensate.} arXiv:math-ph/0410005.


\bibitem{EY}
L. Erd\H os and H.-T. Yau: {\sl Derivation of the nonlinear
Schr\"odinger equation from a many-body Coulom system} Adv. Theor.
Math. Phys. {\bf 5} \, (6) \, (2001), 1169--1205.

\bibitem{FL}
J. Fr\"ohlich and E. Lenzmann: {\sl Mean field limit of quantum
Bose gases and nonlinear Hartree equation.} Preprint
arXiv:math-ph/0409019.

\bibitem{GV} J. Ginibre and G. Velo: {\sl The classical
field limit of scattering theory for non-relativistic many-boson
systems. I and II.} Commun. Math. Phys. {\bf 66}, 37--76 (1979)
and {\bf 68}, 45-68 (1979).

\bibitem{Hepp} K. Hepp: {\sl The classical limit for quantum mechanical
correlation functions. \/} Commun. Math. Phys. {\bf 35}, 265--277
(1974).

\bibitem{He}
I. Herbst: {\sl Spectral theory of the operator
$(p\sp{2}+m\sp{2})\sp{1/2}-Ze\sp{2}/r$.} Comm. Math. Phys. {\bf
53} \, (3) \, (1977), 285--294.

\bibitem{L} E. Lenzmann. In preparation.

\bibitem{LY}
H. Lieb and H.-T. Yau: {\sl The {C}handrasekhar theory of stellar
collapse as the limit of quantum mechanics.} Comm. Math. Phys.
{\bf 112} \, (1) \, (1987), 147--174.

\bibitem{RS}
M. Reed and B. Simon: Methods of mathematical physics. Vol. I.
Academic Press, 1975.


\bibitem{Ru} W. Rudin: Functional analysis.
McGraw-Hill Series in Higher Mathematics, McGraw-Hill Book~Co.,
New York, 1973.

\bibitem{Sp} H. Spohn:
{\sl Kinetic Equations from Hamiltonian Dynamics.}
    Rev. Mod. Phys. {\bf 52} no. 3 (1980), 569--615.
\end{document}